\documentclass[iop]{emulateapj}

\usepackage{apjfonts}
\usepackage{pbox}
\usepackage{bm}
\usepackage{CJK}
\usepackage[breaklinks,colorlinks,citecolor=blue,linkcolor=magenta]{hyperref}
\usepackage{soul}

\newcommand{\set}[1]{\bm{#1}}
\newcommand{\starlabel}{\ell}
\newcommand{\labels}{\set{\starlabel}}
\newcommand{\given}{\,|\,}

\shorttitle{Accelerated Fitting of Stellar Spectra}
\shortauthors{Ting, Conroy, \& Rix}

\begin{document}

\begin{CJK*}{UTF8}{gbsn}
\title{Accelerated Fitting of Stellar Spectra}
\author{Yuan-Sen Ting (丁源森) \altaffilmark{1}, Charlie Conroy\altaffilmark{1}, Hans-Walter Rix\altaffilmark{2}}
\altaffiltext{1}{Harvard--Smithsonian Center for Astrophysics, 60 Garden Street, Cambridge, MA 02138, USA}
\altaffiltext{2}{Max Planck Institute for Astronomy, K\"onigstuhl 17, D-69117 Heidelberg, Germany}

\slugcomment{Submitted to ApJ}

%
%
%
%
%
%
\begin{abstract}
Stellar spectra are often modeled and fit by interpolating within a rectilinear grid of synthetic spectra to derive the stars' labels: stellar parameters and elemental abundances. However, the number of synthetic spectra needed for a rectilinear grid grows exponentially with the label space dimensions, precluding the simultaneous and self-consistent fitting of more than a few elemental abundances. Shortcuts such as fitting subsets of labels separately can introduce unknown systematics and do not produce correct error covariances in the derived labels. In this paper we present a new approach -- {\sc chat} (Convex Hull Adaptive Tessellation) -- which includes several new ideas for inexpensively generating a sufficient stellar synthetic library, using linear algebra and the concept of an adaptive, data-driven grid. A convex hull approximates the region where the data lie in the label space. A variety of tests with mock datasets demonstrate that {\sc chat} can reduce the number of required synthetic model calculations by three orders of magnitude in an 8D label space. The reduction will be even larger for higher-dimensional label spaces. In {\sc chat} the computational effort increases only linearly with the number of labels that are fit simultaneously. Around each of these grid points in label space an approximate synthetic spectrum can be generated through linear expansion using a set of ``gradient spectra'' that represent flux derivatives at every wavelength point with respect to all labels. These techniques provide new opportunities to fit the full stellar spectra from large surveys with $15-30$ labels simultaneously.
\end{abstract}


%
%
%
%
%
%

\section{Introduction}
\label{sec:introduction}

Despite many decades of research, many aspects of Milky Way evolution \citep[see review from][]{rix13} and the Local Group galaxies \citep[e.g.,][]{kir14,wei15} remain unsettled. To unravel the formation history of the Milky Way, spectroscopic surveys are currently being carried out to gather elemental abundances and kinematic information of stars across the Galaxy. High-resolution spectra of $10^5- 10^6$ stars are being collected through surveys such as APOGEE \citep{maj15}, GALAH \citep{des15} and Gaia-ESO \citep{smi14} and with the resolution power of $R \simeq 20,000$ and S/N $\simeq 100$. With the exquisite spectra of these stars, the goal is to measure $15-30$ elemental abundances of each star as precise as possible. Since most stars are long lived and the elemental abundances of galaxies built up gradually over time, these abundances are tell-tale signs of the Milky Way's evolution. Furthermore, stars that formed together are believed to share exceptionally similar elemental abundances \citep[e.g.,][]{des07b,tin12b,fri14,bov15}. By looking for stars that share similar abundances, one goal in these surveys is to reconstruct star clusters that are now disrupted and dispersed in the Milky Way \citep[e.g.,][]{lad03,kop10,dal15}, an idea commonly known as chemical tagging \citep{fre02,bla10a,bla10b,tin15a}. 
 
Identifying members of disrupted star cluster is an important missing piece to understanding the Milky Way. Stars are believed to have migrated from their birth orbit since they formed either through ``radial migration'' \citep[see observational evidence from][]{loe11,kor15} or ``blurring'' of orbits (e.g., through n-body scattering). For example, it has been proposed that stars could radially migrate when corotating with transient structures such as the Galactic bar and spiral arms \citep{min10,ros12,dim13,hal15}. But quantitative, direct observational evidence for radial migration remains scarce, and chemical tagging can provide it.

Chemically tagging disrupted star clusters also informs us about the past star cluster mass function \citep[e.g.,][]{tin15b}. This information is crucial as studies \citep[e.g.,][]{esc08,krui12} have shown that the maximum aggregate size that star formed might depend on the star formation rate and the gas mass in the past. Since most of the clusters are soon disrupted \citep{lad03} constraining the past cluster mass function through chemical tagging may be the best option \citep{bla15,tin15b} .

But all this requires efficient and precise abundance determinations from vast sets of observed spectra. This is challenging for two reasons. First, synthetic spectra for large surveys have systematic uncertainties, as 1D models in local thermodynamic equilibrium (LTE) are typically used for generating a synthetic spectral library \citep[e.g.,][]{smi14,gar15}. Studies have shown that, at least for metal-poor stars, 3D non-LTE calculations are essential for accurate recovery of labels \citep[e.g.,][]{ber15}. Second, generating synthetic spectra is computationally expensive. Even for 1D-LTE models, each synthetic spectrum can take hours to generate, which renders the generation of a synthetic library with $15-30$ elemental abundances impossible with the rectilinear grid approach.\footnote{A rectilinear grid is a model grid that has uniform spacing with a fixed interval for each label.}

In this paper, we will tackle the second challenge by presenting {\sc chat} (Convex Hull Adaptive Tessellation), a set of techniques for fitting stellar spectra by generating a synthetic library using the idea of an adaptive grid and a convex hull. Our method reduces the complicated interpolation-minimization process into a simple series of linear regressions. In \S\ref{sec:limitation} we will discuss the limitations of the rectilinear grid approach. In \S\ref{sec:methodology} we describe the idea and the implementation of our method. We present a comparison of our method with the rectilinear grid approach in \S\ref{sec:results} and show that our proposed method here can reduce the number of models by three orders of magnitude in an 8D label space, and the reduction will be more significant at a higher dimensional label space. This method opens up new possibilities to perform an ab-initio fitting of observed spectra with more labels. We explore some of these possibilities in \S\ref{sec:results}, and we conclude in \S\ref{sec:conclusions}. We emphasize that these techniques can be used for fitting any set of synthetic spectra to observations. In this paper we focus on 1D LTE models but note that as 3D non-LTE models become computationally affordable, {\sc chat} can be applied to those models as well. 

In this paper we focus on techniques directly applicable to automated pipelines for large surveys in which the full spectrum (or portions thereof) are fit to models. However, many of the techniques discussed here are also applicable to the classical technique of fitting equivalent widths of selected spectral features.

%
%
%
%
%
%

\section{The Rectilinear Grid \& Its Limitations}
\label{sec:limitation}

For a rectilinear grid, the number of models grows exponentially with the number of dimensions. For example, in the case of the APOGEE Survey \citep{hol15}, the rectilinear grid is comprised of 6 main labels including $T_{\rm eff}$, $\log g$, the overall metallicity [$Z$/H], the $\alpha$-enhancement [$\alpha$/$Z$], [C/$Z$] and [N/$Z$]. A mere five grid points per dimension would require $5^6 \simeq 15,000$ models. Therefore, any additional dimension such as microturbulence, $v_{\rm turb}$, stellar rotation $v \sin i$ or additional abundances [$X$/H], are very computationally expensive to include. Consequently, the full observed spectra are in practice only fit with these few main labels. Other elements are individually determined in a second step by fitting narrow spectroscopic windows and assuming a fixed underlying atmosphere. But this two-step approach entails several potential problems:

\begin{enumerate}
\item As we show in Appendix~\ref{sec:full-consistent-calculations}, a wide array of elements impact the atmospheric structure. By only considering a subset of important elements when computing grids of atmospheres, one introduces biases in the final spectra. This issue is more apparent for the low-$T_{\rm eff}$ stars (e.g., below $\sim4000$ K). Therefore, the two-step approach could introduce non-negligible systematic biases when determining the photosphere structure by only fitting the main labels.

\item Fitting the full spectra with only a few basic labels (e.g., $T_{\rm eff}$, $\log g$, [$Z$/H] and [$\alpha$/$Z$]) requires assumptions on how the other (not fit) elements trace those labels. A common assumption is that all $\alpha$-elements trace each other, and all other elements trace [Fe/H]. Although this is a good working assumption, without which the fitting would be much worse, this assumption is not true in detail. Therefore, this simplification incurs systematic offsets in the determinations of the main labels such as $T_{\rm eff}$, $\log g$ and $v_{\rm turb}$. 

\item By fitting elemental abundances one at a time, with fixed atmosphere structure, one cannot evaluate their covariances with other labels. For example, elemental abundance determinations depend on $T_{\rm eff}$, and therefore, abundances must be correlated to some level. The covariance matrix is crucial for any chemical tagging studies. Although stars that formed together are believed to be homogeneous to the level of $0.05$ dex \citep[e.g.,][]{des07b,tin12b,fri14} or better \citep{bov15,liu16}, the spread of a cluster in elemental abundances space is dominated by the measurement uncertainties. Therefore, to look for overdensities, we would need to evaluate this spread of a chemical homogeneous cluster in the elemental abundances space \citep{lin13}. \citet{tin15b} showed that ignoring the covariance matrix and using only the marginal uncertainty of each element will increase the background contamination in a search for overdensities by a factor of $10^4$ in a 10D space. But any estimate of the covariance matrix requires that all labels are fit simultaneously, again incurring prohibitive computational expense with rectilinear grid fitting.

\item Restricting the fits of individual elemental abundances to ``clean'' narrow spectral windows excludes other spectral information such as blended lines. From information theory, one can calculate the theoretically achievable precision with the Cramer-Rao bound \citep{cra45,rao45}. If we have accurate synthetic spectra and an effective way to fit all elemental abundances simultaneously using the full spectrum, as illustrated in Appendix~\ref{sec:CR-bound}, we could in principle achieve a precision of $\sim 0.01$ dex for APOGEE data if the systematic errors in the models are smaller than this limit. For comparison, the windows currently defined by the APOGEE pipeline only exploit $\sim 10\%$ of the spectral information, implying that abundance precision will be $\sqrt{10} \sim 3$ times worse than the formal limit (see Appendix~\ref{sec:CR-bound} for details). We note that these ideal theoretical precisions might not be achievable due to systematic uncertainties in the models, but we should still expect a decent improvement in the precision as we are using more information in the spectra.

\item To improve the interpolation within a rectilinear grid, sophisticated algorithms can be employed. For example, in the case of the APOGEE Survey, the interpolation is done with a cubic B\'ezier function. For this algorithm, many spectra from the rectilinear grid are required for each iteration. This implementation is memory intensive, to the extent that not all wavelength points from each spectrum are saved \citep{gar15}. Certain compressions are needed, and some information within the spectra is unavoidably discarded in the compression process. As we will discuss in more details in \S\ref{sec:methodology}, the method proposed in this paper reduces the complicated interpolation-minimization process into a simple series of linear regressions. This method is extremely memory effective, and no compression of spectra is needed.
\end{enumerate}

%
%
%
%
%
%

\section{A New Approach}
\label{sec:methodology}
 
To overcome the challenges discussed in \S\ref{sec:limitation}, we propose a new approach to fitting stellar spectra with ab-initio models, which we call {\sc chat} (Convex Hull Adaptive Tessellation) that has two central elements: (a) around any point in label space for which we have calculated a synthetic spectrum, there exists a (high-dimensional) hypersphere in label space within which the spectrum (i.e., the flux at every wavelength point of a normalized spectrum) varies {\it linearly} with changes in any of the labels; as this region is defined through its series expansion, we refer to this region as a ``Taylor-sphere.''\footnote{Strictly speaking, since we are looking for the best linear interpolations given fixed end points, a more appropriate nomenclature should be ``Legendre-sphere", because (a) we are searching for the best multivariate Legendre polynomial approximations to the first order and (b) we are not calculating the gradient spectra with infinitesimal changes in label space.} The number of synthetic spectral models needed to describe any spectrum that lies within the Taylor-sphere of a model grid point grows linearly with the dimensionality of label space, instead of exponentially. (b) Rectilinear grids in high-dimensional spaces are highly inefficient in covering high-dimensional distributions, especially ones that are as correlated and irregular as the distribution of stars in abundance space. Here we develop a data-driven approach to finding a near-minimal set of grid points whose surrounding Taylor-spheres cover all of the relevant label space. Obviously, this requires at least some a priori knowledge of the distribution of stars in the label space. In addition, {\sc chat} simplifies the interpolation-minimization spectral fitting process to a series of linear regression problems around a manageable set of grid points. In \S\ref{sec:basic}, we will expand the basic ideas of the method. In \S\ref{sec:attractive}, we discuss some of the attractive properties of this method compared to the rectilinear grid approach, and we move on to the implementation details in \S\ref{sec:implementation}.

%
%
%
%
%
%

\subsection{Basic concepts}
\label{sec:basic}

\subsubsection{Gradient spectra and Taylor-spheres}

The predicted continuum-normalized flux of a spectrum, $f_{\rm model}(\lambda|\labels )$ of a model that is specified by a set of labels, $\labels$, changes from point to point in that space, but does so ``smoothly'', or differentiably. The spectrum corresponding to any $\labels$ sufficiently close to a model grid point $\labels_*$ can therefore be described with sufficient accuracy by 
\begin{equation}
f^{\rm lin}_{\rm model}(\lambda|\labels_* + \bold{\Delta}\labels) \simeq f_{\rm model}(\lambda|\labels_*) + 
\overrightarrow{\nabla}_{\labels} f_{\rm model} (\lambda|\labels_* ) \cdot \bold{\Delta}\labels .
\end{equation}

\noindent
In $N_{\labels}$-dimensional label space, the calculation of $\overrightarrow{\nabla}_{\labels}$ requires the calculation of (only) $N_{\labels}$ additional model spectra, used to define the vector of ``gradient spectra'', $\overrightarrow{\nabla}_{\labels} f_{\rm model} (\lambda|\labels_* )$. In this study, we evaluate gradient spectra in finite differences. For each dimension, we derive the 1D gradient spectrum via ${\rm d} f_{\rm model} / {\rm d} \starlabel = (f(\starlabel_1)-f(\starlabel_2))/(\starlabel_1-\starlabel_2)$. The assumption of linearity implies that gradient spectra are decoupled from one another, and the variation of a spectrum from a label point to another can be approximated by the sum of variation in each dimension. This (only) linear scaling of model numbers with $N_{\labels}$ is one of the key advantages to {\sc chat}. The region in label space for which this $1^{st}$-order Taylor expansion is a sufficiently good approximation, we call a ``Taylor-sphere''. For labels $\labels$ that lie within the Taylor-sphere of $\labels_*$ spectral fitting then becomes a simple regression.

\citet{nes15a} show that even for the important labels such as [Fe/H] and $T_{\rm eff}$ that plausibly have small Taylor-radii, the spectral variation at all wavelength points across the entire label space for all giants as a function of the labels can be approximated by a quadratic polynomial function. Therefore, we expect the Taylor-radii to cover finite, and not terribly small regions of the relevant label range. In principle, the high-dimensional linear expansion of the space of model spectra could be extended to a general polynomial expansion. Considering 2$^{nd}$-order expansion, this would require $\sim N_{\labels}$ times more ab-initio model calculations, but may dramatically increase the size of the corresponding ``2$^{nd}$-order Taylor-sphere'', as suggested by the empirical success of the Cannon \citep{nes15a}, but we will leave this exploration to future work.

\begin{figure}
\centering
\includegraphics[width=0.45\textwidth]{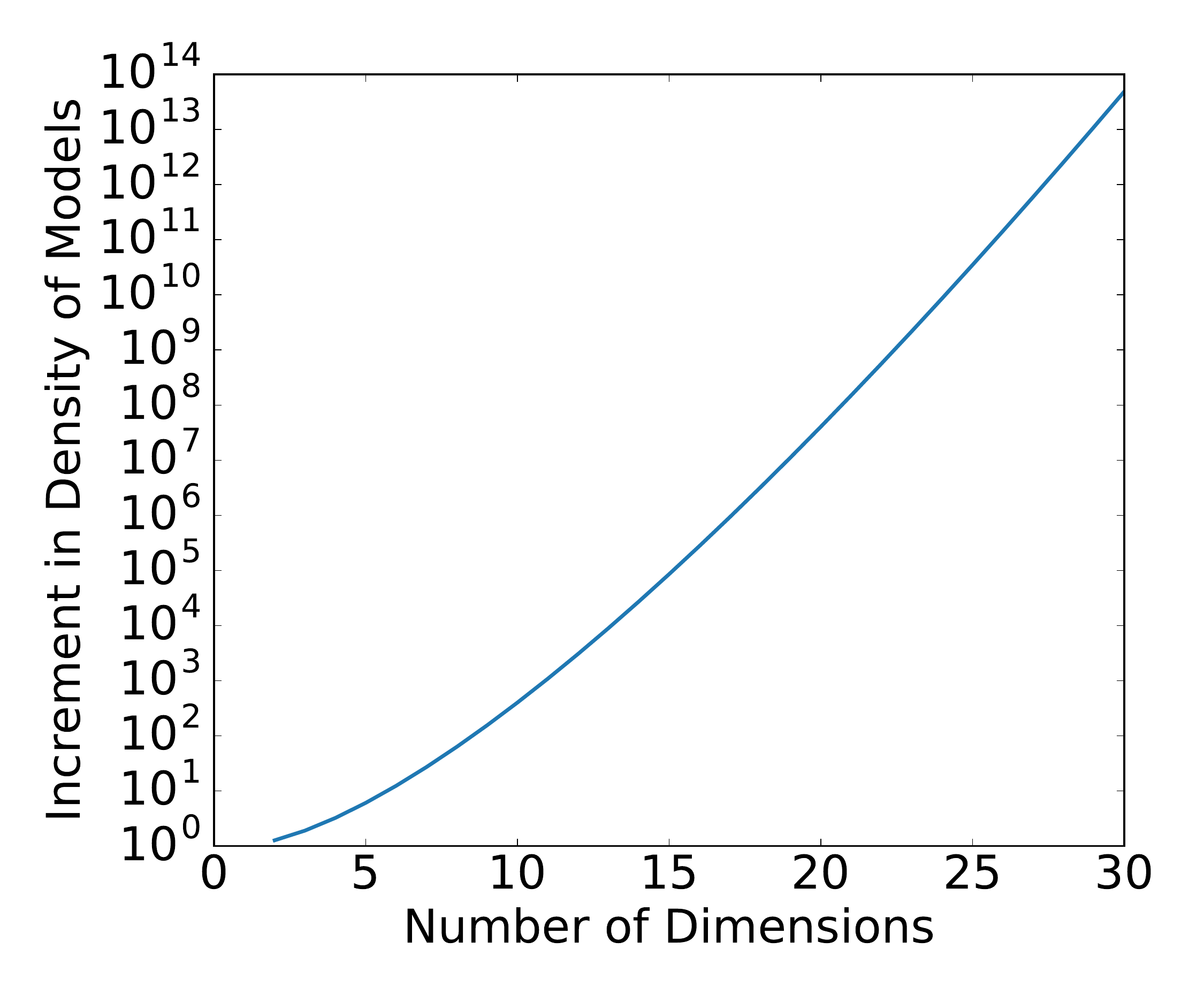}
\caption{The volume ratio of an N-dimensional unit hypercube to a unit hypersphere. This figure shows that using a convex hull significantly improves the density of models in the label space. For example, if the stellar properties only lie in a unit hypersphere in a 10D label space, a rectilinear grid will need $500$ times more models than a convex hull approach that only generates models within the hypersphere. If we consider a label space of 30D, the ratio increases to $5 \times 10^{13}$. In practice, since stellar elemental abundances only live in a $7-9$ dimensional subspace in a 30D space, which is more compact than a hypersphere, the improvement is more significant than the ratio shown in this figure.}
\label{fig:volume-ratio}
\end{figure}

\begin{table*}
\begin{center}
\caption{Comparisons of the rectilinear grid approach and {\sc chat}.\label{table:compare}}
\begin{tabular}{lcc}
\tableline \tableline
\\[-0.2cm]
Properties & Rectilinear grid & {\sc chat}
\\[0.1cm]
\tableline
\\[0.0cm]
Defining label space & \begin{tabular}{c} Results from previous (smaller) \\ surveys are not required \end{tabular} & Uses previous results to determine where models are needed \\[0.35cm]
Number of models & \begin{tabular}{c} Grows exponentially with \\ the number of dimensions \end{tabular} & \begin{tabular}{c}Grows linearly with the number of dimensions in the limit \\ where spectra vary linearly with respect to the additional dimension\end{tabular} \\[0.35cm]
Number of dimensions & Limited to $\lesssim$\,7D label space & \begin{tabular}{c}Can extend to >10D label space and allows us to perform ab-initio \\ full spectral fitting for multiple elements\end{tabular} \\[0.35cm]
Fitting procedure & \begin{tabular}{c}Requires computationally intensive \\ interpolation-minimization algorithms\end{tabular} & \begin{tabular}{c} Reduces the problem to a series of simple linear regressions \\ For an 8D space, we reduce the computational time by 100 folds \\ when compared to a quadratic interpolation \end{tabular} \\[0.5cm]
Data-driven models & Not compatible with data-driven methods & \begin{tabular}{c}Since the fitting relies on gradient spectra, the synthetic gradient spectra \\ can potentially be substituted with data-driven empirical gradient spectra \end{tabular} \\[0.35cm]
Abundance precision & \begin{tabular}{c}Limited recovery of the labels due \\ to the sparse density of models. For an 8D \\ space, we would need at least $\mathcal{O}(10^5)$ models \end{tabular} & \begin{tabular}{c} Better recovery because the density of models in the label space \\ increases. We achieve the same precision with $\mathcal{O}(10^2)$ models \end{tabular} \\[0.6cm] 
\tableline
\end{tabular}
\end{center}
\end{table*}

%
%
%
%
%
%

\subsubsection{A data-driven model grid}

The next step in {\sc chat} is to find a ${\rm (near-)}$minimal set of grid points, $\labels_*$, so that the ensemble of their surrounding Taylor-spheres covers all the relevant label space. If we had such a set of $\labels_*$, and if that set was manageable small, then the entire fitting procedure would be reduced to a set of linear regressions.

To start this, we illustrate concretely how important it is to abandon rectilinear grids in high dimensions. The key point is that the volumes of hyperspheres of unit radius differ drastically from the volumes of hypercubes with unit length in high-dimensional space. Fig.~\ref{fig:volume-ratio} illustrates this volume ratio: for a 10-dimensional space, the volume ratio of the hypercube over the hypersphere is $\sim 5 \times 10^2$; and in a 30-dimensional space, the ratio increases to $\sim 5 \times 10^{13}$! Therefore, if we place our synthetic models (at $\labels_*$), only in the (hyper)spherical region where they are needed (rather than in an encompassing hypercube), the density of models grows exponentially with the number of dimensions, for a given total number of models. Furthermore, \citet{tin12a} showed that, in a $\sim 25$-dimensional elemental abundances space, stellar elemental abundances are contained in a 7-9 dimensional subspace, the volume of which is necessarily smaller than the $25$-dimensional hypersphere. Therefore, the gain will be much more significant in practice.

Clearly, we only need to calculate spectral models and their gradient spectra in regions of label space containing data. To define such a region, we use the concept of a convex hull. A convex hull is the minimal convex polygon that encompasses all the data points. Of course, for any given survey we do not know which part of label space the data cover. But there is sufficient information from existing surveys that one can define an approximate convex hull \citep[e.g.][]{ben14,hol15}. In detail, one would want to carefully consider the construction of the convex hull for the specific problem of interest. 

\begin{figure*}
\centering
\includegraphics[width=0.9\textwidth]{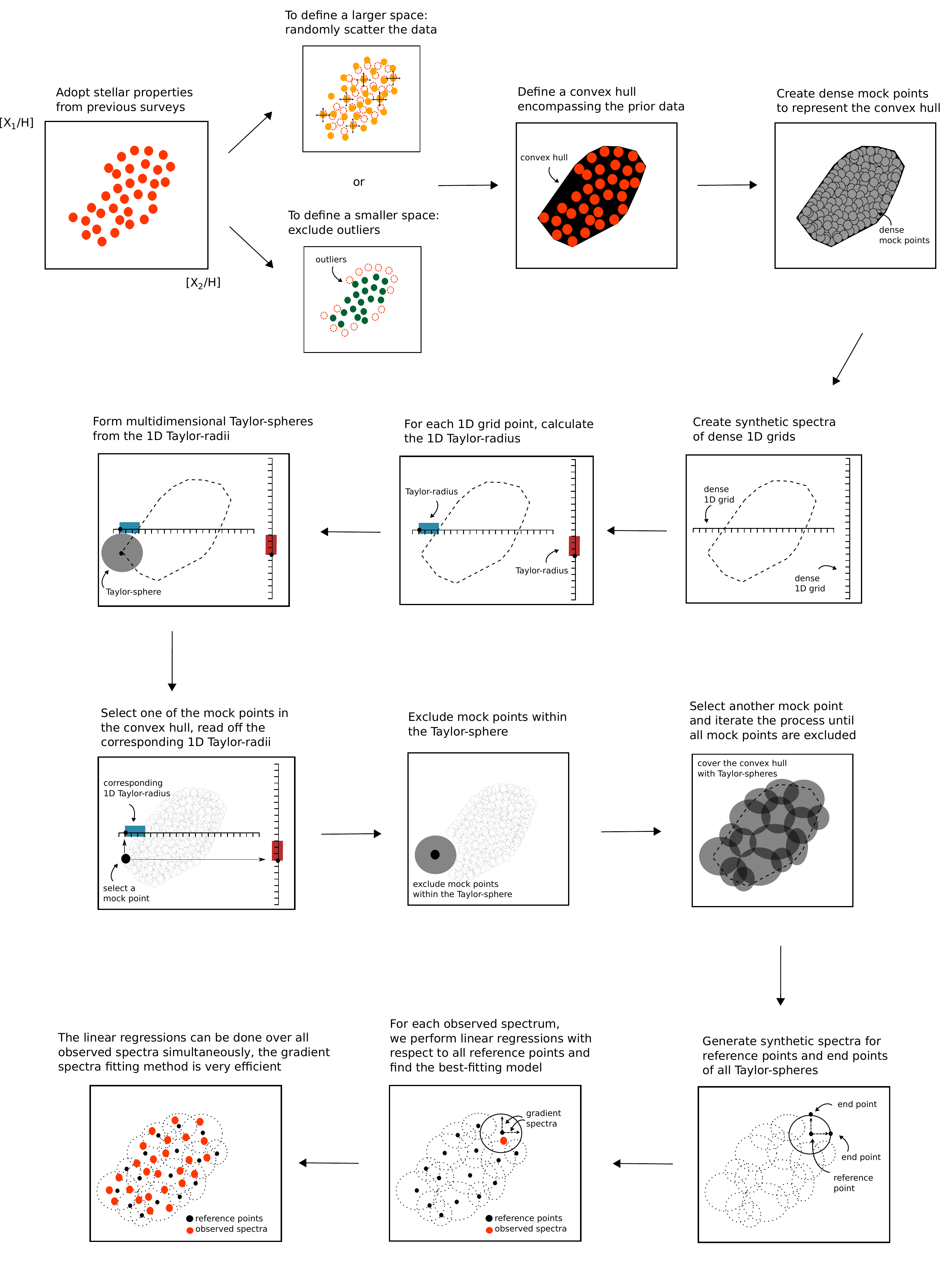}
\caption{A schematic illustration of {\sc chat} and its implementation. In this illustration, we demonstrate a 2D scenario. In practice, we generalize this approach to $10-30$-dimensional label space using the same idea.}
\label{fig:illustration}
\end{figure*}

After this region is defined, we need to find the set of $N_{\rm mod}$ grid points at $\labels_*$ whose surrounding Taylor-spheres cover this minimal polygon. Our approach is illustrated in Fig.~\ref{fig:illustration}. The Figure illustrates how we define the convex hull, find Taylor-spheres that cover the convex hull and reduce the spectral fitting problem to a series of linear regressions. The details are described in \S\ref{sec:implementation}, but the remarkable result is that the number of necessary grid points $N_{\rm mod}$ remains manageable even for a high-dimensional label space.

%
%
%
%
%
%

\subsection{{\sc chat}'s advantages}
\label{sec:attractive}

{\sc chat} has a number of very attractive properties compared to the rectilinear grid model that we summarize in Table~\ref{table:compare}. First, within the Taylor-sphere of a model grid point, $\labels_*$, the computational expense of fitting only grows linearly with the number of labels to be fit. If $N_{\rm mod}$ model grid points are needed to cover the complex hull with their Taylor-spheres, then we only need to calculate $N_{\rm mod} \times (N_{\labels} + 1)$ synthetic models, i.e., for each model grid point, we need to calculate the spectrum at this grid point and additional $N_{\labels}$ models to calculate the array of gradient spectra. As $N_{\labels}$ increases by one, the size of the array of gradient spectra only increases by one, even though we might need slightly more models $N_{\rm mod}$. This slow growth is different from the rectilinear grid approach, where the number of models always grows as $d^{N_{\labels}}$, where $d$ is the number of grid points in each dimension regardless whether or not the spectrum varies linearly with a label.

Second, decomposing the label space into a series of Taylor-spheres reduces the complicated interpolation-minimization spectral fitting process to a series of linear regressions. This calculation can be easily done on any personal computer after the synthetic library is generated and can be easily parallelized. At the same time, we also circumvent the computational memory problem as discussed in \S\ref{sec:limitation}. As we only need to perform linear regression from each grid point separately, only the gradient spectra of a particular grid point are loaded into the memory each time. The memory requirements for {\sc chat} are very modest.

Aside from the computational speedup, the use of Taylor-spheres to determine where to create models has the important conceptual advantage that the step size in each dimension is determined in a statistically-rigorous manner (such that the error induced by assuming linear interpolation is below a predetermined tolerance).

Fitting models with linear regressions through gradient spectra also provides a natural connection to fully data-driven technniques. In this paper, we calculate the gradient spectra according to synthetic models. But one could replace the theoretical gradient spectra with empirical gradient spectra if we have enough training set to span the label space. As discussed in \citet{nes15a}, performing spectral fitting in a fully data-driven way is advantageous in some cases because such an approach produces elemental abundances on the same overall scale set by the training set.

%
%
%
%
%
%

\subsection{Implementation}
\label{sec:implementation}

With the qualitative picture of Fig.~\ref{fig:illustration} in mind, we now describe some of the specifics of finding suitably sized Taylor-spheres and filling the data convex hull with these Taylor-spheres.

%
%
%
%
%
%

\subsubsection{Defining the Taylor-spheres}
\label{sec:define-Taylors}

So far we have defined the Taylor-spheres qualitatively as the regions around a spectral model at $\labels_*$ within which the vector of gradient spectra describes all model spectral ``sufficiently'' well. We now describe the quantitative procedure, which starts by determining the 1D Taylor-radii in all the label directions. We generate a fine 1D grid for each of these dimensions with a step size of $\Delta$[$X$/H]$\,= 0.03\,$dex, $\Delta T_{\rm eff} = 25\,$K, $\Delta \log g = 0.05$, $\Delta v_{\rm turb} = 0.05\,$km/s. When evaluating the Taylor-radius in any one label-space coordinate, we adopt fiducial labels for the other coordinates. Focusing here on APOGEE red clump stars, we choose $T_{\rm eff} = 4750\,$K, $\log g = 2.5$, $v_{\rm turb} = 2\,$km/s and Solar metallicity. We checked that the Taylor-radii are relatively insensitive to the choice of the fiducial values for the other coordinates.

To evaluate the Taylor-radius for any one dimension of label space, {\em for each grid point $\starlabel_*$}, we use the finely-spaced 1D grid to find the maximum distance from the grid point in label space such that all points, $\Delta\starlabel$ in between the grid point and the end point $\Delta\starlabel_{\rm max}$ can be well approximated by linearly interpolating between these two points; the quality of the approximation is judged by the $\chi^2$ difference between the interpolated spectrum and the directly calculated spectrum at that label point. We define this $\chi^2$ to be
\begin{eqnarray}
\label{eq:chi2-criteria}
\chi^2(\starlabel_*, \Delta\starlabel , \Delta\starlabel_{\rm max}) \qquad \qquad \qquad \qquad \qquad \qquad \qquad \nonumber \\
\equiv
\sum_{\lambda} \frac{\Bigl(f_{\rm interp}(\lambda\given \starlabel_*, \Delta\starlabel , \Delta\starlabel_{\rm max})-f_{\rm model}(\lambda\given \starlabel_*, \Delta\starlabel)\Bigr )}{\sigma_\lambda^2},
\end{eqnarray}

\noindent
where we sum over all the uncorrelated wavelength points, each of which has a typical error of $\sigma_\lambda$ in the actual data set. We deem a spectrum to be well enough interpolated if $\chi^2 < \epsilon$, where $\epsilon$ is the tolerance that we set. If we were to set $\epsilon$ to be the number of fitted labels, the uncertainties of the estimated labels due to interpolation errors would be comparable to uncertainties due to observation noise. Since we typically fit $\mathcal{O}(10)$ labels, we choose $\epsilon \simeq 50$ (assuming S/N=100) \footnote{We could have chosen $\epsilon = 10$, but this requires more models in our test cases. Since the purpose of this paper is to compare {\sc chat} with the rectilinear grid approach, we decided to reduce the computational time by adopting a slightly higher $\epsilon$. As long as we adopt the same number of models for both cases, the comparison is fair.} in our study. We also verified that the systematic uncertainties in elemental abundances from interpolation errors (of the order $0.01$ dex, see plots in \S~\ref{sec:rectilinear-vs-adaptive}) are indeed comparable to the one due to observation noise (Appendix~\ref{sec:CR-bound}, assuming S/N$\,=100$), justifying this choice. Finally, we define the Taylor-radius of any grid point $\starlabel_*$ as the maximum $\Delta\starlabel_{\rm max}$ for which $\chi^2(\starlabel_*, \Delta \starlabel , \Delta\starlabel_{\rm max})\le \epsilon $, for every grid point $\Delta\starlabel \; {\rm where} \; 0 \le \Delta\starlabel\le \Delta\starlabel_{\rm max}$. We found that $T_{\rm eff}$ has a Taylor-radius of $175\,$K for $T_{\rm eff} = 4000\,$K and a Taylor-radius $350\,$K for $T_{\rm eff} = 5000 \,$K; $\log g$ has a $\sim$ constant Taylor-radius of $0.9\,$dex for $\log g = 1-5$; [Fe/H] (an element that has a lot of absorption lines) has a Taylor-radius of $0.5\,$dex in the low metallicity regime [Fe/H]$\,=-1$, and the Taylor-radius decreases to $0.3\,$dex for solar metallicity; [K/H] (a trace element) has a Taylor-radius of $2\,$dex for [K/H]$\,=-1$, and $1\,$dex for solar metallicity.

We have thus far defined the 1D Taylor-radii, which results in a Taylor-sphere encompassing all points $\labels$ with
\begin{equation}
\sum_{n=1}^{N_{\labels}}
\Bigl (
\frac{\starlabel_n - \starlabel_{*,n}}{\Delta\starlabel_{{\rm max},n}}
\Bigr )^2 \le 1,
\end{equation}

There is no guarantee that all points within this Taylor-sphere satisfy the linearity conditions only tested along the coordinate axes. We have tested this issue in several ways. In 2D we verified this assumption by studying grids of random element pairs. For each element pair, we first found the 1D Taylor-radii set by a fixed $\chi^2$ criterion. We found that 2D grid points within an ellipse defined by the 1D Taylor-radii fulfill the same criterion, and grid points outside the ellipse violate the criterion. The ellipse-approximation holds very well for all element pairs that are not strongly correlated in the spectra space. For strongly coupled elements, we performed a few tests with CNO and found that the $\chi^2$ contours appear to be more irregular and do not form a perfect ellipses, even though the simple ellipse approximation still performs reasonably well. We will leave this issue to be explored in more detail in future work.

\begin{figure*}
\centering
\includegraphics[width=0.9\textwidth]{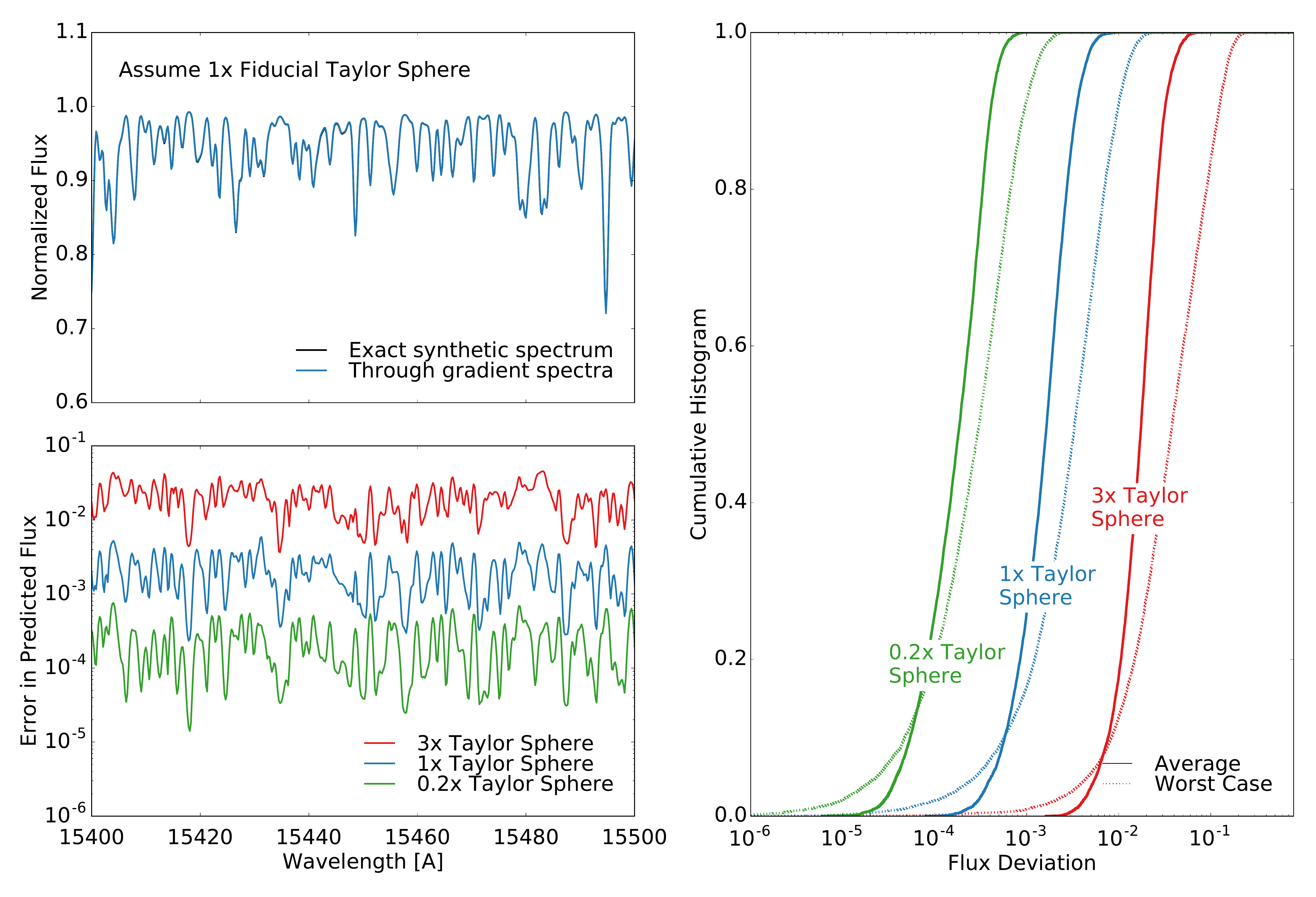}
\caption{Demonstration of the fidelity of synthetic spectrum reconstruction through gradient spectra. We vary all 15 elements in APOGEE and the three stellar parameters, $T_{\rm eff}$, $\log g$ and $v_{\rm turb}$. We consider a $\chi^2$ tolerance $\epsilon = 50$, the tolerance we assume in this study, and evaluate the Taylor-radii and their corresponding Taylor-sphere. The Taylor-radii are evaluated at the fiducial reference point of $T_{\rm eff} = 4750\,$K, $\log g = 2.5$, $v_{\rm turb} = 2.0\,$km/s and Solar metallicity. To illustrate the behavior as a function of Taylor-radii, we shrink the radii 5 times and expand $3$ times which we denote as $0.2 \times\,$ Taylor-sphere and $3 \times\,$Taylor-sphere, respectively. We generate 100 mock spectra for labels that are within each of these Taylor-spheres and compare the gradient spectra reconstructions to the ab-initio calculations. The top left panel shows the comparison of a segment averaging all 100 trials. The comparison is excellent, justifying our choice of $\chi^2$ tolerance. The bottom left panel shows the mean absolute differences between the exact and gradient-interpolated models assuming three different Taylor-spheres. In the right panel we plot the cumulative histogram of the wavelength-by-wavelength deviations. The solid lines demonstrates the absolute deviations averaged over all 100 trials, and the dashed lines show the worst-case scenarios of the 100 trials. For the fiducial Taylor-sphere, the wavelength-by-wavelength deviation is well below the typical APOGEE S/N of 100.}
\label{fig:demonstrate}
\end{figure*}

In Figure~\ref{fig:demonstrate}, we demonstrate how well the gradient spectra method works by comparing the gradient spectra reconstructions within a Taylor-sphere and the ab-initio model calculations. We calculate the Taylor-sphere centered at a fiducial reference point of $T_{\rm eff} = 4750\,$K, $\log g = 2.5$, $v_{\rm turb} = 2.0\,$km/s and Solar metallicity. We draw random labels within the Taylor-sphere and calculate their ab-initio model spectra. The gradient spectra constructions are done by taking the reference spectrum at the reference point and linearly adding the gradient spectra multiplied by the step size in each label dimension. We vary all 15 elements in APOGEE and the three main stellar parameters, $T_{\rm eff}$, $\log g$ and $v_{\rm turb}$. We show that for a tolerance of $\epsilon = 50$, the wavelength-by-wavelength deviation is comfortably below the typical APOGEE S/N of 100. As we will demonstrate in \S\ref{sec:results}, we only need a few of these Taylor-spheres to span the relevant label space, which implies that through {\sc chat}, we can reconstruct synthetic spectra with a near-minimal number of synthetic spectra calculations.

%
%
%
%
%
%

\subsubsection{Filling the convex hull of label space}

We now discuss how to find the set of model grid points, $\labels_*$, so that their surrounding Taylor-spheres fill the convex hull of the pertinent portion of label space. For a high dimensional space, data points that use to determine the convex hull (the minimum polygon) might only cover a small fraction of the volume. In order to make sure that we are covering the convex hull sufficiently well with Taylor-spheres, we start by representing the convex hull by $10^6$ uniform mock data points. As we will end up with $<1000$ model grid points (see \S\ref{sec:results}), $10^6$ mock points will sample the convex hull sufficiently well for our purposes. We construct the convex hull using {\sc convexhull} routine in the {\sc scipy.spatial} package. To determine whether a point is within the convex hull, we use {\sc delaunay} routine and check whether the point is within a simplex of the Delaunay tessellation.

Following Figure \ref{fig:illustration} we then successively identify model grid points, $\labels_*$. We start by picking a random point from the mock data within the convex hull. We read off their 1D Taylor-radii corresponding to the closest grid points in the finely-spaced 1D grids. Then we ask which of our mock points fall within the Taylor-sphere determined by these 1D Taylor-radii, and eliminate those. Then we draw another random point from the {\it remaining} mock points. We take it to be the next model grid point, consider its surrounding Taylor-sphere, and remove the mock points that lie within that sphere. We emphasize that the Taylor-radii read off from the 1D grids for each random point are difference -- we assume $\starlabel_*$ that is closest to the random point when evaluating the Taylor-radii. This ``adaptive'' approach takes into account that the radii could vary, for example, in a low $T_{\rm eff}$ regime compared to a high $T_{\rm eff}$ regime. We repeat this procedure until no mock points remain. The resulting set of $\labels_*$ is the near-minimal set of model grid points, whose Taylor-spheres fill the convex hull. 

We refer to this set of grid points as a {\em near}-minimal set since there are overlaps between Taylor-spheres, so it might not be the absolute minimal set. However, the overlaps of hyperspheres are quite small in a higher dimensional space. For an 8D label space that we will explore in \S~\ref{sec:rectilinear-vs-adaptive}, through Monte Carlo integration we found that the total volume of all the ellipsoids within the convex hull is only $\sim 5$ times the volume of the convex hull. So at most, we can only reduce the number of grid points by another factor of five.

This procedure requires no calculation of any model spectra apart from the precalculated spectra of the fine 1D grids in determining the Taylor-radii. After we have the set of $\labels_*$, we calculate the model spectra at each of them, along with the $N_{\labels}$ gradient spectra at each $\labels_*$. These are all the models needed for {\sc chat}.

There are also a number of practical choices to be made to define the convex hull in the first place. If the prior information used to determine the convex hull is very noisy or has outliers, the volume of the convex hull will be larger than the intrinsic volume within which the labels of the sample at hand reside \citep[e.g.,][]{tin15b}. With many dimensions, the volume of the noisy convex hull could be much larger than the intrinsic volume, ``wasting'' many model grid points at the periphery of the volume. To address this problem, one could run a kernel density estimation to map out the density of the label space, similar to \citet{tin15b} and cull outliers according to the density map. If, however, a goal of the spectral fitting is to look for rare outliers in label space, one might consider randomly scattering the prior data so that they define a larger volume in label space (see Figure \ref{fig:illustration}). A point to keep in mind is that the particular implementation of {\sc chat} will depend on the problem of interest.

\begin{table*}
\begin{center}
\caption{Nomenclature of element classification in this paper.\label{table:element}}
\begin{tabular}{lcc}
\tableline \tableline
\\[-0.2cm]
Nomenclature & Element & Classification
 \\[0.1cm]
\tableline
\\[0.0cm]
Primary elements & Fe, C, N, Mg, Si & Used in atmosphere calculations\footnote{These elements have important influences to the atmosphere. See Appendix~\ref{sec:full-consistent-calculations}.} \\[0.1cm]
Secondary elements & O, Na, Al, S, K, Ca, Ti, V, Mn, Ni & Held fixed at solar metallicity or trace primary elements when creating model atmospheres \\[0.1cm]
Trace elements & \multicolumn{2}{c}{Elements other than the 15 elements in APOGEE DR12 and should have negligible effect on the atmosphere}
\\[0.2cm] 
\tableline
\end{tabular}
\end{center}
\end{table*}

%
%
%
%
%
%

\subsection{Creating stellar models}

The model atmospheres and spectra for this work are computed with the {\sc atlas12} and {\sc synthe} programs written and maintained by R. Kurucz \citep{kur70,kur81,kur93}. We adopt the latest line lists provided by R. Kurucz\footnote{\href{http://kurucz.harvard.edu}{\tt http://kurucz.harvard.edu}}, including line lists for TiO and H$_2$O, amongst many other molecules. Model atmospheres are computed at 80 zones down to a Rosseland optical depth of $10^3$, and each model is automatically inspected for numerical convergence. We adopt the \citet{asp09} solar abundance scale. Convection is modeled according to the standard mixing length theory with a mixing length of 1.25 and no overshooting. Spectra are computed with the {\sc synthe} program and are sampled at a resolution of R$\,=300,000$ and then convolved to lower resolution.

%
%
%
%
%
%

\subsection{Spectral fitting}

After calculating all $N_{\rm mod}$ spectra at the model grid points $\labels_*$, along with models needed to evaluate their gradient spectra, we have reduced the spectral fitting of one object to a set of $N_{\rm mod}$ linear regressions. In practice it may be more advantageous to go to one model grid point, $\labels_*$ and execute a trial linear regression fit on all spectra of the sample. In the end the best fitting labels for each object will be the fit (among the $N_{\rm mod}$) that has the lowest $\chi^2$. 

Let us now denote the entire (observed) spectrum of the $n^{th}$ $\big ( n\in [1,N_{\rm sample}]\big )$ object as the vector $\bold{f}_{{\rm obs},n}$ of length $N_{\rm pix}$, with the vector of its uncertainties denoted as $\bold{e}_{n}$, where $N_{\rm pix}$ is the number of uncorrelated wavelength points in the spectrum. The entire model spectrum at a grid point $\labels_*$ can be denoted as the equally long vector $\bold{f_{\rm mod}}(\labels_* )$. The set of $N_{\labels_*}$ gradient spectra associated with $\bold{f_{\rm mod}}(\labels_* )$ can be denoted as a $N_{\rm pix}\times N_{\labels_*}$ matrix $\bold{GL}(\labels_*)$. For linear regression, this defines a covariance matrix $\bm{\Omega}_n(\labels_*)$:
\begin{eqnarray}
\bm{\Omega}_n(\labels_*) = \bold{GL}^{\rm T}(\labels_*)\cdot \Big(\bold{GL}^{\rm T}(\labels_*) \times \bold{e}_n^{-2} \Big)^{\rm T}.
\end{eqnarray} 
Then the best fitting labels $\bm{\labels_n}$, based on the model grid point $\labels_*$
are given by
\begin{eqnarray}
\bm{\labels_n}\big (\labels_* \given {{\{\rm data\}}_n}\big ) \qquad \qquad \qquad \qquad \qquad \qquad \qquad \qquad \nonumber \\
\ \ \ \ \ \ \ =\bm{\Omega}_n^{-1} (\labels_*)\cdot \Biggl ( \bold{GL}^{\rm T} (\labels_*)\cdot \Big(\big (\bold{f}_{{\rm obs},n}-\bold{f_{\rm mod}}(\labels_* )\big ) \times \bold{e}_n^{-2}\Big) \Biggr ),
\end{eqnarray} 

\noindent
where ${\rm data}_{n} \equiv [\bold{f}_{{\rm obs},n},\bold{e}_{n}]$.

This linear regression fitting process is extremely efficient even though it is performed $N_{\rm mod}$ times because the regression can be done analytically and consumes little memory even for multiple variables. Also the main computational cost in spectral fitting lies in generating a model grid instead of the fitting process, and for the former, the number of models in {\sc chat} grows much more benignly with more dimensions than an exponential growth. Furthermore, linear regressions around different $\labels_*$ can also be parallelized to speed up the process. Finally, in this regime we have the covariance matrix $\bm{\Omega}_n$ analytically. This covariance matrix, reflecting the label space error ellipsoid, is a critical component in chemical tagging studies \citep[see][]{tin15b}.

%
%
%
%
%
%

\section{Results and implications}
\label{sec:results}

We are now in a position to explore what {\sc chat} can deliver in practice. We will cover three related aspects: in \S\ref{sec:rectilinear-vs-adaptive} we explore how well {\sc chat} (with $N_{\rm mod}$ grid points, and hence a total of $N_{\rm mod} \times (N_{\labels} + 1)$ models) does in determining large numbers of labels from data, compared to a sensibly chosen, but rectilinear high-dimensional grid that has a total of $\sim N_{\rm mod} \times (N_{\labels} + 1)$ grid points. We also explore what systematic errors can occur when fitting a higher dimensional spectra with a lower dimensional effective label space (\S\ref{sec:other-elements}). In \S\ref{sec:pca} we explore how to reduce the effective dimensionality of label space, exploiting {\it astrophysical} correlations among elemental abundances through a principal component analysis (PCA). Finally, we discuss some limitations of {\sc chat} in \S\ref{sec:adaptive-limitations}.

Throughout these tests we use mock spectral data drawn from synthetic spectral models that are based on up to 18 labels: three stellar parameters $T_{\rm eff}$, $\log g$, $v_{\rm turb}$ and up to 15 APOGEE elemental abundances; the remaining elements, which we will call ``trace elements'', are assumed to scale with [Fe/H] at solar abundance ratios. In other words, we assume labels (from the APOGEE DR12 red clump stars), create {\sc atlas12} models with these labels and use these models as our mock data to be fit. That means that there always exists a set of labels that leaves no systematic differences between the data and the model. We defer the application of {\sc chat} to real data to another paper.

\begin{figure*}
\centering
\includegraphics[width=1.0\textwidth]{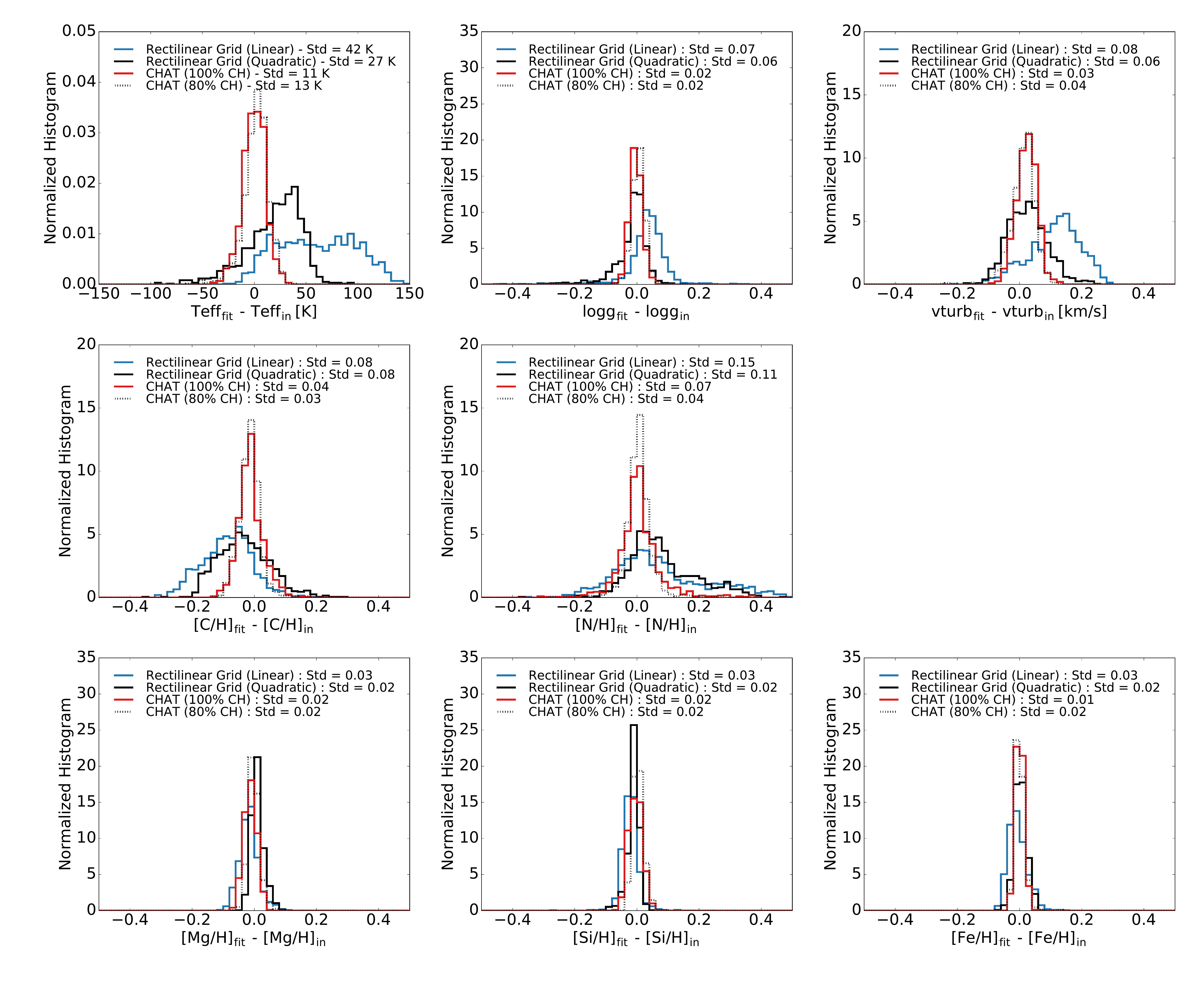}
\caption{Recovery of labels for 1000 synthetic test models with APOGEE red clump properties. We consider 8D test models, assuming solar abundances for the other elements, and fit the test models with 8D synthetic libraries. The blue lines show the results of a linearly interpolated rectilinear grid with $\sim 6500$ grid points. The solid black lines show the results from the same rectilinear grid but the grid is now quadratic B\'ezier interpolated. To enable a fair comparison with {\sc chat}, the rectilinear grid only spans the range of label values used in this test. The red lines show the results of the standard approach of {\sc chat} with the same number of models as in the rectilinear grid. Labels are recovered very well, in all cases more precisely than when using the rectilinear grid. But {\sc chat} can be further improved by excluding outlying points before defining the convex hull. If we exclude the $20\%$ most outlying points, as shown in the dotted black lines, we can further reduce the number of models needed by a factor of $\sim 35$ and achieve the same precision. In this case, only $180$ models are needed in an APOGEE red clump synthetic library.}
\label{fig:fit-8D-8D}
\end{figure*}

\begin{figure*}
\centering
\includegraphics[width=1.0\textwidth]{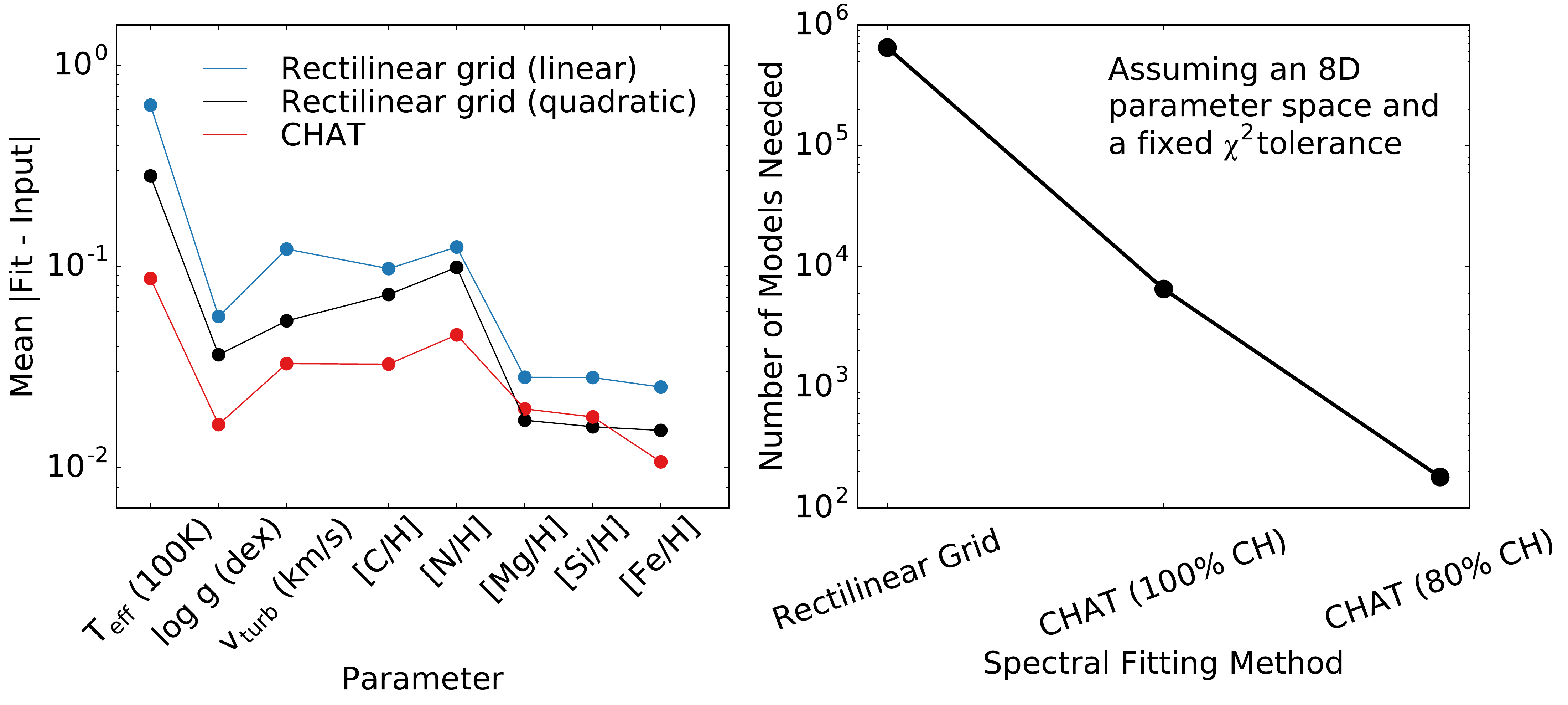}
\caption{A summary of Fig.~\ref{fig:fit-8D-8D} and a comparison of {\sc chat} with the rectilinear grid approach. The left panel shows the mean deviation of the synthetic spectra input recoveries in an 8D label space. In all three cases, we consider $\sim 6500$ models. A quadratic interpolation improves the recoveries from a rectilinear grid, but it is still less precise than {\sc chat} and is about $10^2$ times slower than {\sc chat}. The right panel shows the estimated number of models needed to have the same recovery precision as {\sc chat}. To have the same density of models in the label space, the rectilinear grid approach requires two orders of magnitude more models, i.e., $6.5 \times 10^5$ models. {\sc chat} can be further improved by discarding the 20\% outlying data when determining the convex hull. In this case, we can further reduce the number of models by at least another order of magnitude. Only $180$ models are needed for the improved {\sc chat}.}
\label{fig:summary-fit-8D-8D}
\end{figure*}

\begin{figure*}
\centering
\includegraphics[width=1.0\textwidth]{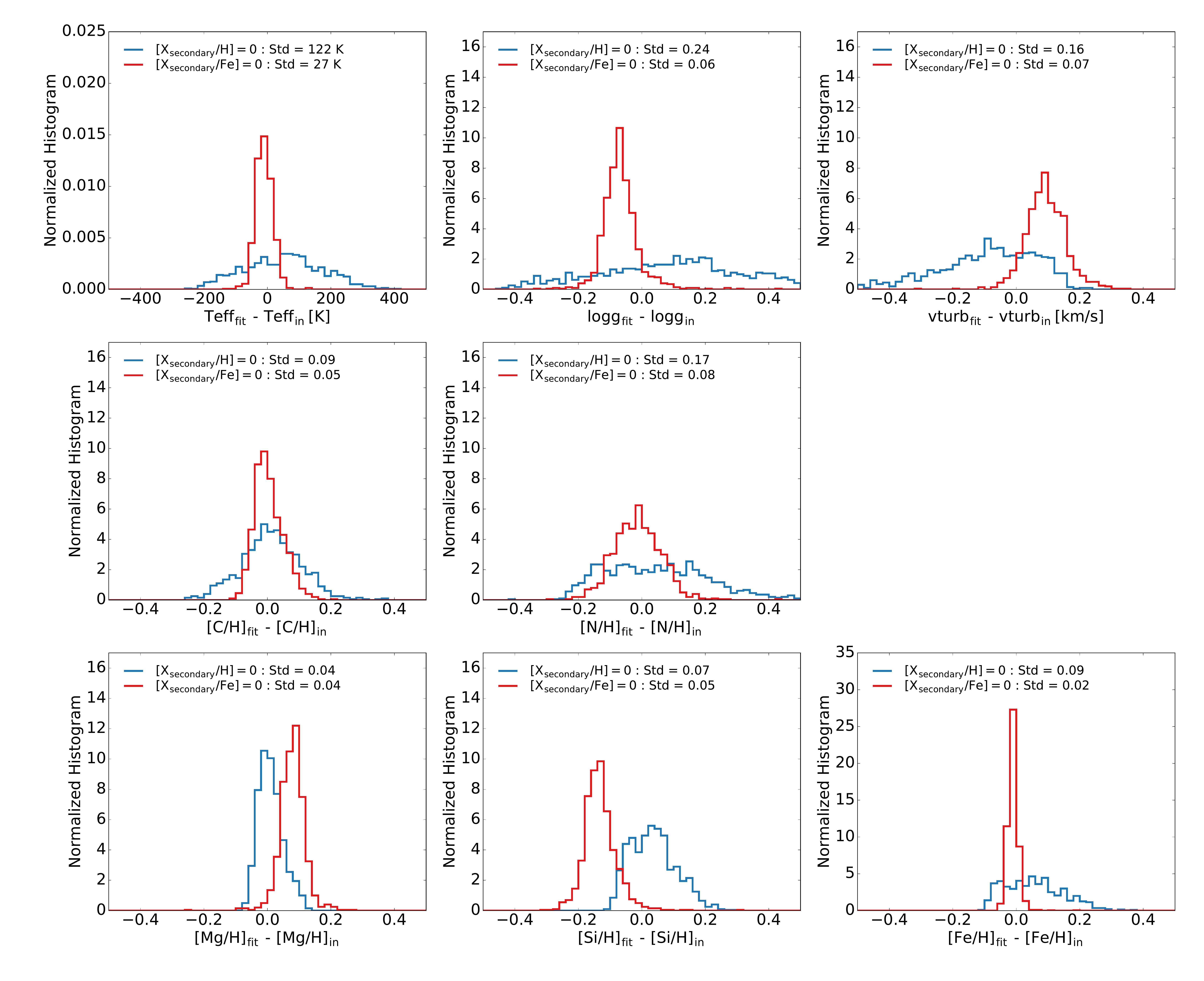}
\caption{Similar to Fig.~\ref{fig:fit-8D-8D}, but here we consider 1000 synthetic test models that vary all 15 elements, $T_{\rm eff}$, $\log g$ and $v_{\rm turb}$, whose labels are drawn from 1000 APOGEE red clump stars. We fit these test models with lower-dimensional synthetic libraries. In the first case, shown in the blue lines, we consider the $\sim 6500$ models 8D {\sc chat} library in Fig.~\ref{fig:fit-8D-8D}, fixing the other elements to be solar metallicity. In this case, the fit is unsatisfactory, showing that some assumptions have to be made to approximate high-dimensional spectra with lower-dimensional synthetic libraries. In the second case, shown in the red lines, we assume that the $\alpha$-elements trace each other and the other elements trace [Fe/H]. We generate the same number of models as the previous case. The fits improve significantly as we take into account the other elements beside the fitted labels. However, these assumptions on the other elements are not true in detail, and hence systematic offsets remain in some of the fits.}
\label{fig:fit-8D-18D}
\end{figure*}

\begin{figure*}
\centering
\includegraphics[width=1.0\textwidth]{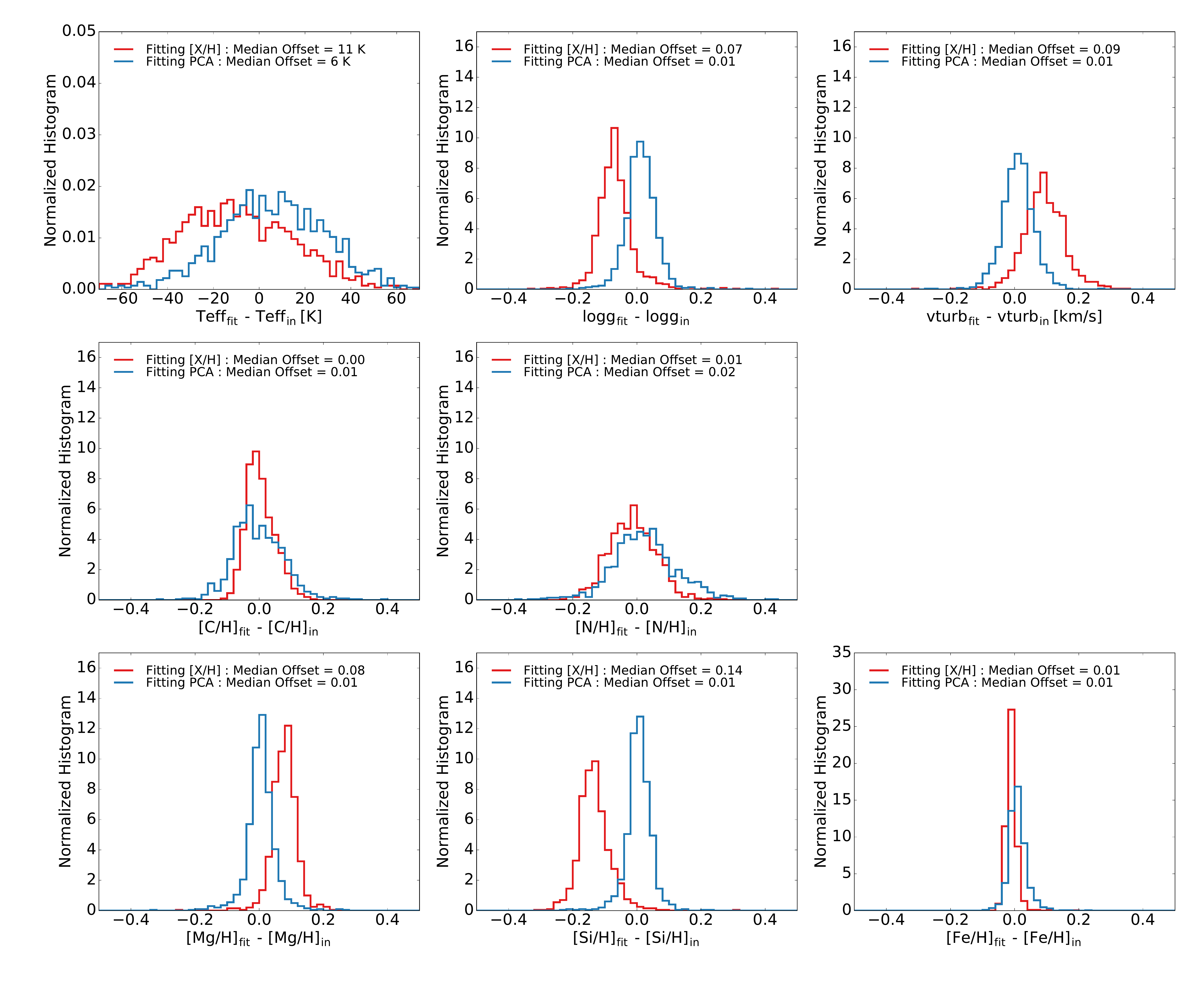}
\caption{An effective way to reduce systematic offsets is to fit the coefficients of PCA components in elemental abundances space instead of fitting [$X$/H]. The red lines show the same results in Fig.~\ref{fig:fit-8D-18D}. We assume that the $\alpha$-elements trace each other and the other elements trace [Fe/H]. Since this assumption is not true in detail, fitting the global metallicity [$Z$/H] and the $\alpha$-enhancement leaves some systematic offsets. If we fit the coefficients of principal components instead, as shown in the blue lines, we reduce the systematic offsets because we take into account the correlations of elements more properly.}
\label{fig:fit-pca-18D}
\end{figure*}

%
%
%
%
%
%

\subsection{Rectilinear grid fitting vs. {\sc chat}}
\label{sec:rectilinear-vs-adaptive}

In this section we compare how well we recover the input labels from mock spectra using two fitting approaches, a rectilinear grid approach and {\sc chat} using the same number of models. For this test case, we consider an 8D label space: three main stellar parameters -- $T_{\rm eff}$, $\log g$, $v_{\rm turb}$ and five main elements, which we will call ``primary elements''. We chose these elements as they either have a significant influence on the overall atmospheric structure ([Fe/H], [Mg/H], [Si/H]\footnote{Another important element is O, but creating a 9D grid is too computationally expensive for the rectilinear grid approach. However, we checked that replacing Si with O does not alter the conclusions in this paper.}, see Appendix~\ref{sec:full-consistent-calculations}) for red clump stars (i.e., $T_{\rm eff} = 4500 - 5000 \, {\rm K}$), or because they have important molecular features in the H-band APOGEE spectra ([C/H], [N/H]). We define 10 additional elements that are derived in APOGEE DR12 to be ``secondary elements'' and assume solar metallicity for all other ``trace elements''. Our adopted element classification nomenclature is summarized in Table~\ref{table:element}. We sample the 8D label space for the mock spectra by simply adopting as input the labels from the APOGEE red clump sample \citep{bov14} and $v_{\rm turb} = 2.478 - 0.325 \log g$ \citep{hol15}. We make the same selection cut as \citet{hay15,tin15b} to cull APOGEE values that are not reliably determined. When generating test data, we add photon noise corresponding to the median wavelength-dependent S/N of the APOGEE red clump spectra.

We assume that the rectilinear grid approach spans the full range of label values in the APOGEE red clump sample, and we take three grid points in each of the eight dimensions, leading to $\sim 6500$ model grid points. To interpolate this rectilinear grid we consider both a linear interpolation and a quadratic B\'ezier interpolation \citep{mes13a}. The interpolation codes -- {\sc ferre}\footnote{\href{http://www.as.utexas.edu/~hebe/ferre/}{\tt http://www.as.utexas.edu/$\sim$hebe/ferre/}} \citep{all06,all14} are adopted from the APOGEE pipeline and provide a direct comparison to the state-of-the-art rectilinear grid approach. {\sc ferre} performs a wavelength-by-wavelength interpolation of the flux and finds the best- itting spectrum through $\chi^2$-minimization. For {\sc chat}, we only generate models that are within the convex hull of the APOGEE red clump sample. As the stellar parameters $T_{\rm eff}$, $\log g$ and $v_{\rm turb}$ reflects the evolutionary state of the star, while elemental abundances reflect the chemical evolution of the Galaxy when the star formed, they should be uncorrelated to first approximation \citep[but see][]{mas15,mar15}. We consider the convex hull of these two groups separately and cross-product their model grid points and gradient spectra. We checked that if we were to consider the 8D convex hull directly, we would further reduce the number of models needed as there are non-trivial correlations between these two groups. We choose to consider these two groups separately to have a more direct comparison with the PCA method in \S\ref{sec:pca}. We adjust the size of the Taylor-spheres, by tweaking the $\epsilon \simeq 50$ tolerance in the $\chi^2$ criterion in Eq.~\ref{eq:chi2-criteria}, such that the convex hull is filled by Taylor-spheres formed from $\sim 6500$ models (including the model grid points and their associated gradient spectra). 

We then determine the best fitting 8D label set for all mock spectra with both methods. Figure ~\ref{fig:fit-8D-8D} shows the comparison of the results. Even in this relatively low-dimensional, 8D label space, {\sc chat} provides better --- higher accuracy and precision --- label recovery than the rectilinear grid with equally many spectral models. The results also imply that, to achieve the same nominal precision, {\sc chat} will require far fewer models than the rectilinear grid approach.

Part of the explanation for this gain is simple: using Monte Carlo integration, we find that the volume of the APOGEE red clump 8D convex hull is $\sim 100$ times smaller than the hypercube spanned by the rectilinear grid. Therefore, we are sampling the label space $\sim 100$ times denser in {\sc chat}. The rectilinear grid approach would require $\sim 650,000$ models in order to produce the same recoveries with linear interpolation. For example, if we were to adopt the standard APOGEE spacing in a rectilinear grid, we would need at least $5^8 = 4 \times 10^5$ models. This is consistent with the conventional wisdom that for rectilinear grids it is difficult to go beyond $6$D-$7$D label space. Quadratic/cubic interpolation slightly reduces the number of models needed to achieve the same precision, but the computational efficiency is compromised. In this study, we did not consider a cubic interpolation because a cubic interpolation in the rectilinear grid approach requires at least four grid points per dimension; it requires a minimum of $4^8= 65,000$ models, which is again significantly more models than {\sc chat}. In our test case, using a single CPU and the same number of models, we found that it takes about $\mathcal{O}(10)$ minutes for {\sc chat} to find the best fitting labels for 1000 spectra, but it takes $\sim \mathcal{O}(1000)$ minutes for a quadratic B\'ezier interpolation, showing the enormous gain of reducing a complicated interpolation-minimization process to a series of simple linear regressions.

Beyond the label space volume difference, the systematic filling of this (smaller) convex hull with adaptive tessellation results in additional gains, i.e., we put more models in regions where linear interpolations fail. Although not shown, to evaluate the contributions from each of these two aspects, we performed tests by only considering convex hull without adaptive tessellation and an adaptive tessellation in a regular label space without the convex hull. We find that the convex hull plays a more important role because it improves the models density in the label space globally. The adaptive tessellation plays a smaller role, but it is important for $T_{\rm eff}$ because $T_{\rm eff}$ shows the most nonlinearity in model variation. Low $T_{\rm eff}$ regimes need more models than the high $T_{\rm eff}$ regimes because the synthetic spectra vary more drastically when molecules start to form at low $T_{\rm eff}$.

Remarkably, {\sc chat} can be even more efficient. As discussed in \S\ref{sec:implementation}, the periphery of any 8D space comprises most of the volume. And, as illustrated in the right panel at the second last row of Fig.~\ref{fig:illustration}, covering the convex hull with Taylor-spheres centered at the periphery points covers a region larger than the convex hull itself. Through Monte Carlo integration, we found that only $\sim 10\%$ of the total volume of the Taylor-spheres is within the convex hull, the rest of $\sim 90\%$ surrounds the convex hull. The outermost points could have already been covered by the Taylor-spheres around points in the interior. Therefore, we eliminate the $20\%$ most outlying points using kernel density estimation before constructing the convex hull around the other 80\%. We choose an 80\% convex hull because we found that the Taylor-spheres centered at points in this smaller convex hull already cover the full convex hull.

The black dotted lines in Fig.~\ref{fig:fit-8D-8D} demonstrate {\sc chat}'s label recovery in this case: we need only 20 model grid points $\labels_*$ and their associated gradient spectra, i.e., a total of 180 spectra (20 model grid points, and $20\times 8$ models to determine the gradient spectra), to fulfill the same $\epsilon-\chi^2$ criterion as before. The resulting label recovery illustrated in the black dotted lines shows that we can achieve the same precision as the standard implementation of {\sc chat} but with 35 times fewer models. The upshot is that by culling 20\% of the outlying points when defining the convex hull, we are able to reduce the number of the required synthetic spectrum calculations to fit red clump stars in 8D label space from $6 \times 10^5$ in a rectilinear grid approach to only $180$ model spectra.

We summarize the comparison between the rectilinear method and {\sc chat} in Fig.~\ref{fig:summary-fit-8D-8D} for the case of 8D fitting. Note that, for a rectilinear grid, even if we consider two grid points per dimension, we will require $2^8 = 256$ models. The gradient fitting approach here surpasses the fundamental limit of a rectilinear grid because in the limit where the spectra vary linearly with all labels and are decoupled from one another, the number of models needed in {\sc chat} grows linearly instead of exponentially. 

For on-going large spectroscopic surveys, full spectral fitting is limited to a subset of ``main'' labels. For example, in APOGEE DR12 \citep{hol15,gar15}, a 6D label space of $T_{\rm eff}$, $\log g$, [$Z$/H], [$\alpha$/$Z$], [C/$Z$], [N/$Z$] was considered. Even a 6D space with five grid points requires $5^6 \sim 15,000$ models. To make the computational consumption more affordable, some important fitting labels were not included, such as $v_{\rm turb}$\footnote{In DR12, APOGEE found a tight $\log g-v_{\rm turb}$ relation. Thus, this relation was assumed to reduce the number of models needed in the synthetic library.} and $v \sin i$. But stellar rotation could play an important role for low-$T_{\rm eff}$ dwarf stars. Omitting them is believed to be the main reason that labels for cool dwarfs were not robust in DR12 \citep{hol15}. We show that even for an 8D space, using {\sc chat} reduces the number of models by a factor of $\sim 1000$. The reduction of models opens up the opportunity to expand many more dimensions and allows $v \sin i$ and $v_{\rm turb}$ to be included more easily.

%
%
%
%
%
%

\vspace{0.5cm} 
\subsection{Consequences of fitting a subset of the label space}
\label{sec:other-elements}

To fully specify an observed spectrum within its (high-S/N) error bars, one may require the specification of several dozens of labels, encompassing the stellar parameters and {\it all} elements that could contribute to the spectrum. However, the rectilinear grid fitting approach is limited to subspaces of much lower dimension. In that case, assumptions have to be made for not-fitted labels. To explore this effect, we create a set of synthetic test spectra that are specified by the 18 labels measured in the APOGEE red clump sample: 15 elements, along with $T_{\rm eff}$, $\log g$ and $v_{\rm turb}$. We assume that all remaining trace elements follow [Fe/H] at solar ratios.

To start, we try to match these mock spectra with {\sc chat}, but fitting only the 8 of the 18 labels, those shown in \S\ref{sec:rectilinear-vs-adaptive}. Fig.~\ref{fig:fit-8D-18D} (blue lines) illustrates that the label recovery is unsatisfactory. This may not be unexpected, as we only vary five elements in the 8D grid, we necessarily mismatch absorption lines from the other elements, held at Solar ratios. As a consequence, the fit of $T_{\rm eff}$ is compromised, which in turn affects the fit of the other labels. One approach to reducing these systematics, while keeping the number of fitted labels low is to exploit the established astrophysical covariances among elemental abundances: we can, e.g., assume that all $\alpha$-elements trace each other, while all other elements scale with the global metallicity, [$Z$/H], an approach followed in the APOGEE pipeline. If we then generate a 7D grid of model spectra, $T_{\rm eff}$, $\log g$, $v_{\rm turb}$, [$Z$/H], [$\alpha$/$Z$], [C/$Z$] and [N/$Z$] using {\sc chat}, and with the same number of models, and use them to fit the 18D mock spectra. As shown in red lines in Figure~\ref{fig:fit-8D-18D} the systematic errors in the label recovery are strongly reduced. This demonstrates that label recoveries can be good with low-dimensional (e.g., 7D) fitting, if astrophysical label-correlations are properly exploited.

But even in this latter case systematic offsets in the label recovery remain: about $10\,$K for $T_{\rm eff}$, $0.1\,$dex for $\log g$ and $0.1\,$km/s for $v_{\rm turb}$. Even though the $\alpha$ elements broadly trace each other, this is is not true in detail. This suggests the need for a more systematic way to reduce the dimensionality of the label space, an issue which we will address in the next section.

%
%
%
%
%
%

\subsection{Fitting PCA components of elemental abundances}
\label{sec:pca}

Principle Component Analysis (PCA) provides a simple and systematic way to characterize the astrophysical correlations among elemental abundances. As such, it may be an effective way to set element ratios that are not directly fitted labels. A detailed exposition of PCA in elemental abundances space is beyond the scope of this paper, but details can be found in \citet{and12} and \citet{tin12a}\footnote{Note that the APOGEE pipeline uses PCA to compactify the spectral space. In our case, we use PCA to compactify the label space of elemental abundances.}. The main points of \citet{tin12a} can be summarized as follows. PCA measures the correlation among elements. Each principal component is a unit vector in the elemental abundances space. The principal components are ordered according to their contributions to the total variances of the data sample. In practice, only the first few principal components are significant and relevant, because additional PCA components are likely dominated by observational noise. \citet{tin12a} showed that for $15-30$ measured elements, not all elements provide independent pieces of information. Elements fall into groups that span a much smaller $7-9$ dimensional subspace. In turn, measuring the $7-9$ dimensional principal components should be sufficient to predict the abundances of all $15-30$ elements.

With this idea in mind, a more effective way to fit high-dimensional spectra with a lower-dimensional effective subspace is to consider the coefficient of each main principal component as a fitting label instead of the usual [$X$/H]. In this case, we should be able to fully characterize an observed spectra with $< 13$ labels (including the stellar parameters, $T_{\rm eff}$, $\log g$, $v_{\rm turb}$ and $v \sin i$), but still account for $30$ elements. A synthetic library covering $13$-dimensional label space is feasible, but only with the advantages provided by {\sc chat} (see \S\ref{sec:rectilinear-vs-adaptive}).

{\sc chat} easily generalizes to the case of fitting PCA components, which are simply taken as the labels in lieu of the direct abundances. In other words, the model grid points and Taylor-spheres are simply determined in the space of stellar parameters and PCA-components. Specifically, We consider an 8D label space comprising $T_{\rm eff}$, $\log g$, $v_{\rm turb}$ and the five most important principal components. As shown in Figure~\ref{fig:fit-pca-18D}, standard deviations of the PCA label recovery (transformed into the space of abundances) are about the same as that of the element label recovery because the underlying model density in the label space does not improve by transforming into the PCA space. However, systematic offsets in the label recovery are dramatically lower when fitting principal components, because we take into account the element correlations more properly.

%
%
%
%
%
%

\subsection{Current limitations and future directions}
\label{sec:adaptive-limitations}

Despite its attractive properties, {\sc chat} also has limitations. {\sc chat} fundamentally relies on a sensible definition of a convex hull in label space. This should be straightforward for the bulk of Milky Way stars, because we have a basic understanding regarding elemental abundance distributions from previous surveys \citep[e.g.,][]{ben14,hol15}. But the construction of a convex hull may be more problematic in other circumstances, such as in the search for extremely metal-poor stars of the Milky Way, or for surveys of other galaxies. In these cases the rectilinear approach or applying {\sc chat} to a large rectangular label space with a lenient tolerance might be applied as a first pass in order to aid in the definition of the convex hull in detail. Because of the use of a convex hull in deciding where to create models, {\sc chat} may not work well as a tool in searching for interesting outliers. In its present form, {\sc chat} is strictly linear within the Taylor-radius of each label, i.e., it treats the gradient spectrum in each label as independent. We have tested in a few cases that this linear approximation is indeed good within the entire Taylor-sphere, but we have not explored this exhaustively. One way to overcome this limitation is to expand {\sc chat} beyond the 1$^{st}$-order Taylor-sphere. A 2$^{nd}$-order Taylor-sphere requires $N_{\labels}$ times more models to define both the gradient matrix and the curvature matrix. The fitting process will be slower as the fitting is no longer a simple linear regression, but it is reasonable to assume that a 2$^{nd}$-order model would cover a much larger label space per Taylor-sphere (see \cite{nes15a}). We are currently exploring this idea and will defer the details to a later paper.

%
%
%
%
%
%

\section{Summary and Conclusions}
\label{sec:conclusions}

Major ongoing and planned research initiatives to unravel the chemodynamical formation history of our Milky Way are based on determining the properties of vast numbers of individual stars. This is to be done by taking high-quality spectra of $10^5-10^6$ stars in the Milky Way, and then deriving from them extensive sets of labels, i.e., stellar parameters and $10-30$ elemental abundances. Rigorous spectral modeling would call for all pertinent labels ($N_{\labels} = 10-15$) that characterize one star to be fit simultaneously to its spectrum. For the large data sets at hand this appears, however, computationally infeasible: using established techniques -- based on rectilinear model grids in label space -- the number of required grid points is prohibitive. 

The established response to this quandary is to only fit a few of the labels (typically $4-6$) simultaneously, and determine the other labels separately on the basis of this initial fit. In this paper we have shown that this short cut leads to important systematic errors, given the high data quality, and we offer a solution: an approach with several new techniques which we call {\sc chat}.

{\sc chat}'s defining ideas and capabilities in determining labels, $\labels$, from spectra can be summarized as follows:

\begin{enumerate}
\item Within a sufficiently small patch around a model grid point $\labels_*$ in label space (a ``Taylor-sphere''), any spectrum defined by its label vector $\labels \equiv \labels_* + \Delta\labels$, can be described by a linear expansion around that grid point, as:
\begin{equation}
f_{\rm model}(\lambda|\labels_*) + 
\overrightarrow{\nabla}_{\labels} f_{\rm model} (\lambda|\labels_* ) \cdot \bold{\Delta}\labels ,
\end{equation}

using the ``gradient spectra'', $\overrightarrow{\nabla}_{\labels} f_{\rm model}$. Within this region of label space, spectral model fitting to data is then reduced to linear regression, which is computationally fast. 

\item Only a tiny fraction of the high-dimensional ($N_{\labels}$) label space of stellar parameters and elemental abundances is occupied by real stars. Given prior information, e.g., from existing surveys, an approximate {\it convex hull} can be constructed for this subspace, of dramatically smaller volume than a rectilinear grid in label space encompassing the relevant label subspace. This much smaller label space volume can then be covered far more densely with synthetic model spectra (for a given computational expense). 

\item We have devised an adaptive tessellation of this convex hull with model grid points, $\labels_*$, so that the set of their surrounding Taylor-spheres completely covers label space within the convex hull. Taken together, these elements of {\sc chat} have reduced the daunting task of full spectral fitting to a) the pre-calculation of the grid points, $\labels_*$ and the extent of their ``Taylor-spheres'', b) the calculation of the model spectra and gradient spectra at this modest number of $\labels_*$, and c) linear regression when actually fitting the spectra. The linear regression also implies that the computational expense for producing the needed synthetic spectra only grows linearly with the dimensionality of label space, $N_{\labels}$, not exponentially.

\item We have tested how well (and how fast) {\sc chat} works in practice by considering the case of an $N_{\labels} = 8$ label fit to APOGEE red clump spectra: we show that the number of required model grid points is reduced from $\mathcal{O}(10^5)$ for rectilinear model grids, to $\mathcal{O}(10^2)$ models for the adaptively tessellated grid within the convex hull, improving the computational cost of generating a synthetic library by three orders of magnitude. We also found that {\sc chat} recovers the best fitting labels $100$ times faster than a quadratic B\'ezier interpolation within a rectilinear grid.

\item The dramatically fewer model points, and only linear growth of computation with $N_{\labels}$ makes full spectral fitting with far larger $N_{\labels}$ feasible. Specifically, it now seems possible to simultaneously fit additional labels such as micro-turbulence and stellar rotation, along with 15 elemental abundances.

\item We showed explicitly that fitting spectra that were drawn from a high-dimensional label space with a much smaller $N_{\labels}$, can lead to important systematic errors, unless optimal assumptions about the non-fitted labels are made. The usual way of fitting [$Z$/H] and [$\alpha$/$Z$], assuming other elements are traced by these two characteristics, works well, but some systematic residual offsets remain. If we assume the principal component coefficients to be the fitted labels, the residual offsets are reduced. Fitting principal component coefficients also provides a natural way to perform chemical tagging in a compactified elemental abundances space.

\item With these new techniques, the hope is that we can improve the precision of elemental abundances in these large surveys. As the full elemental abundances space seems to be well-described by $7-9$ element groups, improving the abundance precision by a factor of two will improve the resolving power of star clusters by a factor of $2^{7-9} = 100-1000$ in elemental abundances space. With such an enormous gain in ``resolution'', we might be able to chemically tag Milky Way stars to their birth origins and provide a completely new view of the evolution of the Milky Way.
\end{enumerate}

%
%
%
%
%
%

\acknowledgments

We thank the anonymous referee for useful comments and careful reading of the manuscript. We thank Bob Kurucz for developing and maintaining programs and databases without which this work would not be possible. We thank David W. Hogg for many insightful discussions and Carlos Allende Prieto for a careful read of the manuscript. YST was supported by NASA Headquarters under the NASA Earth and Space Science Fellowship Program - Grant NNX15AR83H and is grateful to the Max-Planck-Institut f\"{u}r Astronomie and the DFG through the SFB 881 (A3) for their hospitality and financial support during the period in which part of this research was performed. CC acknowledges support from NASA grant NNX13AI46G, NSF grant AST-1313280, and the Packard Foundation. HWR's research contribution is supported by the European Research Council under the European Union's Seventh Framework Programme (FP 7) ERC Grant Agreement n.$\,$[321035]. The computations in this paper were partially run on the Odyssey cluster supported by the FAS Division of Science, Research Computing Group at Harvard University. The computational work also used the Extreme Science and Engineering Discovery Environment (XSEDE), which is supported by National Science Foundation grant number ACI-1053575.

%
%
%
%
%
%

\appendix

%
%
%
%
%
%

\section{Can we ignore non-primary elements in the photospheric atmosphere calculations?}
\label{sec:full-consistent-calculations}

Generating stellar synthetic spectra consists of two parts. The first part is the calculation of the model atmosphere. Given stellar parameters and elemental abundances of a star, the temperature, pressure and electron density as a function of the Rosseland opacity of the photosphere can be calculated by solving a system of differential equations ({\sc atlas12}). After this is done, we then proceed with a radiative transfer code ({\sc synthe}) and generate synthetic spectra by integrating over the photospheric atmosphere. With a restricted range in wavelength, the former step is much more time consuming (for APOGEE, it is $\mathcal{O}(10)$ times slower) than the latter step. Since it is too computationally expensive to generate a rectilinear grid of $15-30$ photospheric atmospheres, the standard approach is to make the assumption that the model atmosphere only depends on a few main labels. When determining the spectral variation for secondary elements, only the radiative transfer step is needed.

\begin{figure}
\centering
\includegraphics[width=1.0\textwidth]{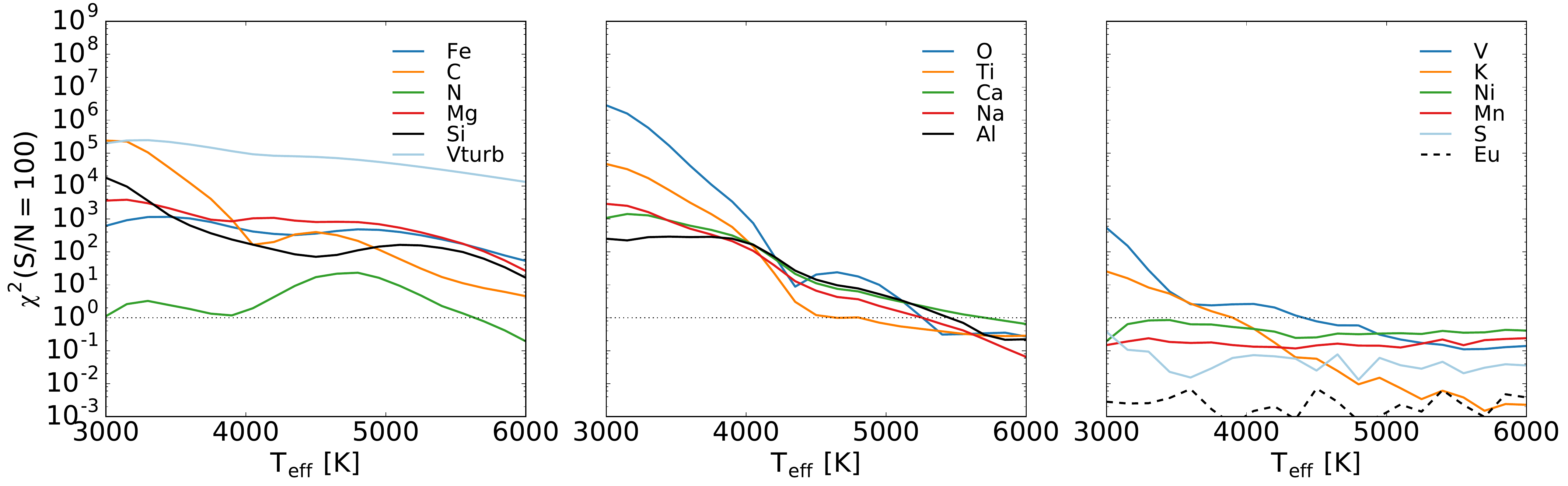}
\caption{Error induced by not creating self-consistent model atmospheres when computing synthetic spectra. We plot the $\chi^2$ between a synthetic spectrum generated with a self-consistent model atmosphere and a spectrum that was computed from a fixed, solar metallicity atmosphere. The underlying reference labels are solar metallicity, $\log g = 2.5$, $v_{\rm turb} = 2\,$km/s, R$\,=20,000$, and S/N$\,=100$ and the wavelength range and bins of the APOGEE survey. We consider a variation of [$X$/H]$\,=0.2$ and $\Delta v_{\rm turb} = 0.5\,$km/s. Results are shown as a function of $T_{\rm eff}$. We do not include predicted lines in this calculation. If the $\chi^2$ values are larger than the number of fitted labels (typically $\mathcal{O}(10)$), the variation is important and distinguishable in the APOGEE survey. The figure shows that many elements, as well as $v_{\rm turb}$, affect the atmosphere substantially, with the largest effects at low temperatures. We also calculate the deviation for Eu as a reference. Eu is a trace element in stars and should have no effect on the atmosphere. Nonetheless, the $\chi^2$ for Eu is not strictly zero ($\sim 10^{-3}$), and we checked that this is due to numerical noise in the atmosphere calculation (the atmosphere is precise to the level 0.1$\,$K for each Rosseland depth layer). Note that $\chi^2$ depends on the S/N quadratically. Hence, assuming $\chi =10$ to be the threshold, we conclude from the Eu result that for spectra with S/N $\lesssim 100 \times \sqrt{10^4} = 10^4$, the numerical noise is negligible for full spectral fitting.}
\label{fig:trace-atmosphere}
\end{figure}

We put this assumption to the test. We consider fully self-consistent calculations for [$X$/H]$\,= 0.2$ dex, varying one element at a time, and compare to the case where we assume a solar atmosphere, ignoring the enhanced contribution from this element. We calculate the $\chi^2$ of these two spectra assuming R$\,=20,000$, S/N$\,=100$ and uncorrelated wavelength points for an APOGEE-like spectrum. In this calculation, we do not include predicted lines (from the atomic line list), in order to be more conservative in our estimate. The real difference could be larger. In Fig.~\ref{fig:trace-atmosphere}, we show that, especially for low-$T_{\rm eff}$ stars, the contributions from secondary elements, as well as $v_{\rm turb}$, can be significant. The differences in these two cases have  $\chi^2$ values larger than the number of fitted labels (typically $\mathcal{O}(10)$). Hence ignoring secondary elements in the atmospheric calculations can bias the abundance determinations for these elements \citep[also read][]{van12}. On the other hand, as shown in the right panel, truly trace elements, such as Eu, indeed have negligible contributions to the atmospheric structure. 

\begin{figure}
\centering
\includegraphics[width=0.8\textwidth]{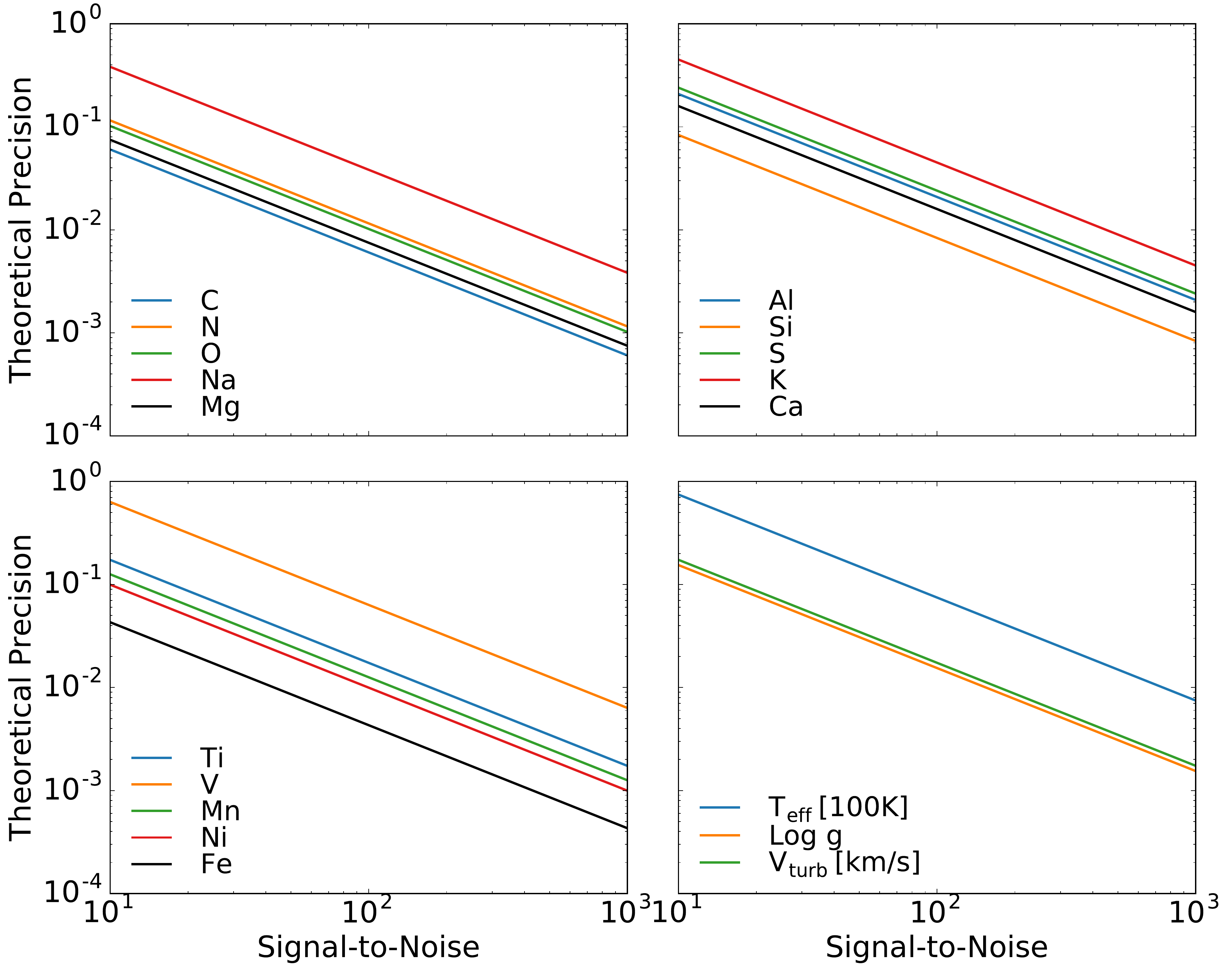}
\caption{Theoretical precision for all 15 elements, $T_{\rm eff}$, $\log g$ and $v_{\rm turb}$ that we could achieve for APOGEE spectra with R$\,= 20,000$ as a function of the spectra S/N. We assume gradient spectra with $\Delta $[$X$/H]$\, = 0.2$, $\Delta T_{\rm eff} = 200\,$K, $\Delta \log g = 0.5$, $\Delta v_{\rm turb} = 0.5$ and with respect to the reference point at solar metallicity, $T_{\rm eff} = 4800\,$K, $\log g = 2.5$, $v_{\rm turb} = 2\,$km/s. We do not include predicted lines in this calculation, and the limit of theoretical precision could be better than what is demonstrated here. If we have robust synthetic models and a way to fit all stellar properties simultaneously, we could, in principle, measure abundances to the precision of $\Delta $[$X$/H]$\sim 0.01$ dex for all 15 elements, $\Delta \log g \sim 0.01$, $\Delta v_{\rm turb} \sim 0.01\,$km/s and $\Delta T_{\rm eff} \sim 10\,$K with S/N$\,=100$ APOGEE spectra.}
\label{fig:CR-bound}
\end{figure}

%
%
%
%
%
%

\section{Theoretical abundance precisions that could be achieved with current on-going surveys}
\label{sec:CR-bound}

Elemental abundances are usually derived from carefully chosen wavelength windows that contain absorption lines that are clean, unblended, and have reliable line parameters (usually calibrated against standards such as Arcturus and the Sun). But more information can be extracted, in principle, if we also consider the blended lines. Furthermore, as illustrated in Appendix~\ref{sec:full-consistent-calculations}, most elements affect the stellar opacity and atmosphere. Therefore, they will indirectly affect the line formation of the other elements. But how well we can extract this indirect information is more questionable than the blended lines. Nonetheless, it might be interesting to understand how much information there is, in principle, in the high-resolution spectra that are currently being collected. We emphasize that these theoretical precisions are not currently achievable (and may never be!) due to systematic uncertainties in the models. Systematic uncertainties aside, the information content depends on three aspects: (a) the number of uncorrelated and independent wavelength points in each spectrum, (b) the extent to which the features, in our case the depths of the absorption lines, vary as a function of the fitted labels, (c) the measurement uncertainty of the normalized flux at each wavelength bin. The measurement uncertainty could be either due to photon noise or imperfect continuum normalization. Here we only consider the ideal case where the uncertainty due to continuum normalization is negligible.

\begin{figure}
\centering
\includegraphics[width=0.6\textwidth]{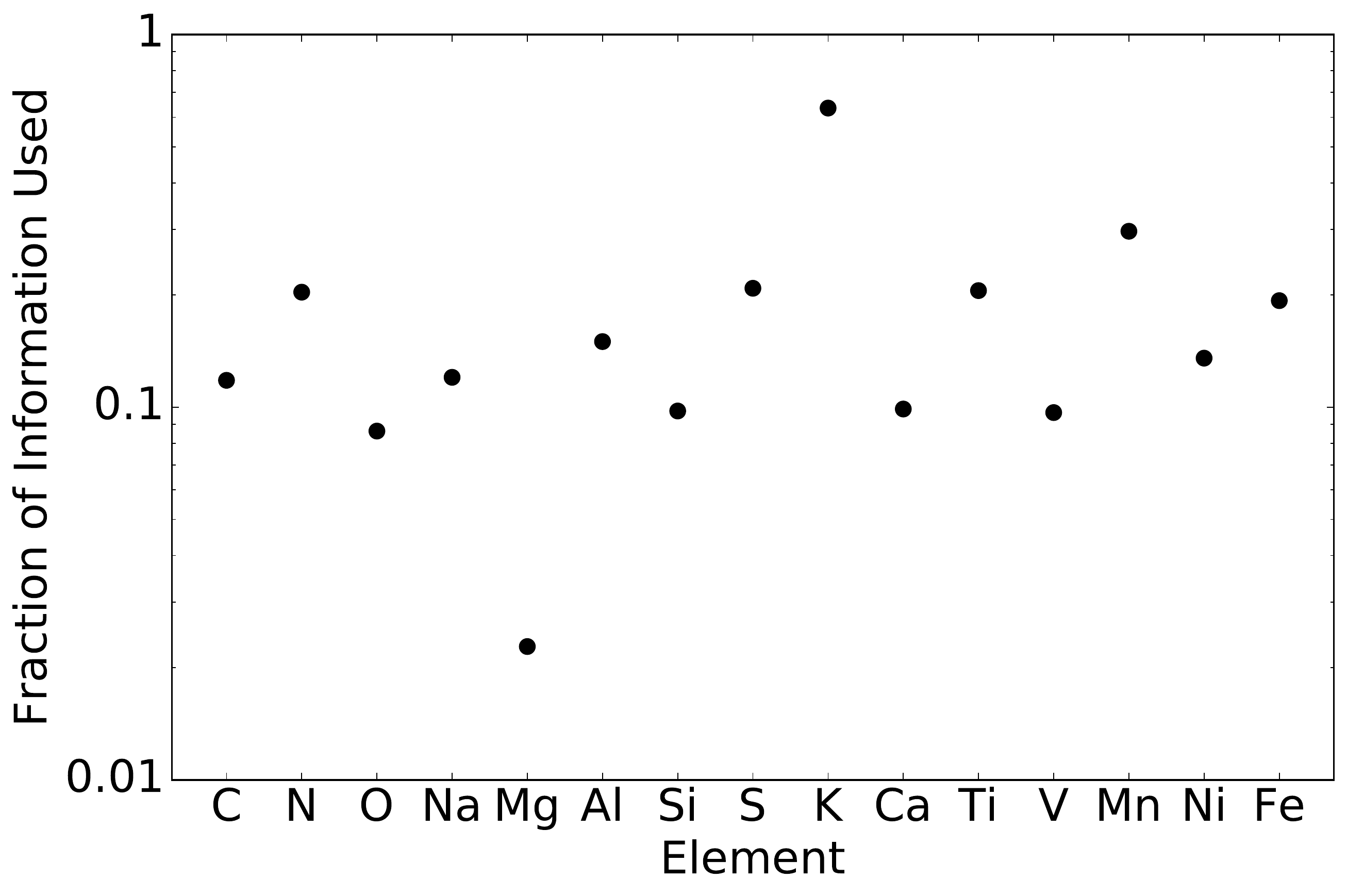}
\caption{Fraction of the total available information used when limiting the fitting to narrow spectroscopic windows compared to fitting the full spectrum. The $y$-axis shows the ratio of the gradient information contained in the APOGEE DR12 spectral fitting windows to the gradient information in the full H-band spectrum (masking regions dominated by telluric absorption or sky lines). For most elements, restricting to narrow windows only exploits $\sim 10\%$ of the total information in the spectra collected by APOGEE. The majority of this extra information is in the form of blended lines, which can be more difficult to interpret.}
\label{fig:missed-information}
\end{figure}

To measure how much spectral features vary as a function of the fitted labels, we consider gradient spectra with $\Delta$[$X$/H]$\,= 0.2$, $\Delta T_{\rm eff} = 200\,$K, $\Delta \log g = 0.5$, $\Delta v_{\rm turb} = 0.5\,$km/s, assuming a reference point at solar metallicity, $T_{\rm eff} = 4800 \, {\rm K}$, $\log g = 2.5$ and $v_{\rm turb} = 2\,$km/s. We generate spectra with R$\,= 20,000$ and in the APOGEE wavelength range. We discard wavelength points that have median uncertainties $> 2\%$ in the real APOGEE spectra because these wavelength points are likely affected by telluric or sky lines. We do not include predicted lines when generating the total gradient information in this calculation to be more conservative in our estimates. We denote the variation in spectrum (the gradient spectrum) to be $\overrightarrow{\nabla}_{\labels} f_{\rm model} (\lambda)_{i}$. For each $(i,\lambda)$, $\overrightarrow{\nabla}_{\labels} f_{\rm model} (\lambda)_{i}$ measures the partial derivative of the absorption line at wavelength $\lambda$ with respect to label $i$. The Cramer-Rao bound \citep{cra45,rao45} predicts that the covariances matrix of the fitted labels, $K_{ij}$ can be calculated from

\begin{equation}
K_{ij}^{-1} = \overrightarrow{\nabla}_{\labels} f_{\rm model} (\lambda_1)_{i} \; C^{-1}_{\lambda_1,\lambda_2} \overrightarrow{\nabla}_{\labels} f_{\rm model} (\lambda_2)_{j},
\label{eq:CR-calculation}
\end{equation}

\noindent
where $C$ is the covariance matrix of the normalized flux. The dot product on the right-hand side serves to sum over the contribution from all wavelength points. For example, we if assume S/N$\,=100$ and only consider uncorrelated wavelength points, we have $C \sim {\rm diag} (10^{-4}, \ldots, 10^{-4})$. The diagonal entries of $K_{ij}$ are the marginalized uncertainties of each label. We plot these marginalized uncertainties for each label as a function of S/N in Fig.~\ref{fig:CR-bound}. The figure shows that for an APOGEE spectrum with S/N$\,=100$, we could achieve a precision of $\sim 0.01$ dex for all elements. 

Finally, Fig.~\ref{fig:missed-information} shows how much information is missed by focusing on narrow spectroscopic windows. The plot shows the ratio of the information content contained in the narrow spectroscopic windows defined and used for abundance measurement by the APOGEE DR12 pipeline to the full spectral range (masking regions dominated by telluric absorption and sky lines). Fig.~\ref{fig:missed-information} shows that, for most elements, the spectroscopic windows misses $\sim 90\%$ of the information. Much of the extra information is contained in blended lines and features that do require accurate models to reliably interpret. As illustrated in Eq.~\ref{eq:CR-calculation}, the measurement precision improves in quadrature with the gradient information. Therefore, we could in principle improve the precision by a factor of three if we can minimize the systematic uncertainties in the models and perform full spectral fitting. This is the task that lies ahead.

%
%
%
%
%
%

\end{CJK*}

\vspace{0.5cm}
\bibliography{biblio.bib}

\end{document}